\documentclass[aps,prl,twocolumn,superscriptaddress,showpacs,preprintnumbers,amssymb,tightenlines,floatfix]{revtex4}
\usepackage[fleqn]{amsmath}
\usepackage{graphicx} 
\usepackage{dcolumn}  
\usepackage{xcolor}
\usepackage{subfigure}
\usepackage{mathtools}
\usepackage{units}
\usepackage{cancel}
\graphicspath{{ps}}
\usepackage{hyperref}
\hypersetup{colorlinks=true}

\begin{document}
%\preprint{\vbox{ \hbox{Belle Preprint 2012-29}
                 %\hbox{KEK Preprint 2012-34}                 
                % \hbox{hep-ex nnnn, if available}
%}}

\title{ \quad\\[0.5cm] Semi-inclusive studies of semileptonic {\boldmath $B_s$} decays at Belle} 

\begin{abstract}
We present an analysis of the semi-inclusive decays $B_s \to D_s^- X \ell^+ \nu$ and $B_s \to D_s^{*-} X \ell^+ \nu$, where $X$ denotes
a final state that may consist of additional hadrons or photons and $\ell$ is an electron or muon. The studied $B_s$ decays are contained in the $121.4~{\rm fb}^{-1}$ $\Upsilon(5S)$ data sample collected by the Belle detector at the KEKB asymmetric-energy $e^+e^-$ collider. The branching fractions of the decays are measured to be $\mathcal{B}(B_s \to D_s^- X \ell^+ \nu) = [8.2 \pm 0.2 (\text{stat}) \pm 0.6 (\text{syst}) \pm 1.4 (\text{ext})]\%$ and $\mathcal{B}(B_s \to D_s^{*-} X \ell^+ \nu) = [5.4 \pm 0.4 (\text{stat}) \pm 0.4 (\text{syst}) \pm 0.9 (\text{ext})]\%$, where the first two uncertainties are statistical and systematic and the last is due to external parameters. The measurement also provides an estimate of the $B_s^{(*)}\bar{B}_s^{(*)}$ production cross-section, $\sigma(e^+e^- \to B_s^{(*)}\bar{B}_s^{(*)}) = (53.8 \pm 1.4 (\text{stat}) \pm 4.0 (\text{syst}) \pm 3.4 (\text{ext}))\,{\rm pb}$, at the center-of-mass energy  $\sqrt{s} = 10.86\,{\rm GeV}$.
\end{abstract}


\noaffiliation
\affiliation{University of the Basque Country UPV/EHU, 48080 Bilbao}
\affiliation{Beihang University, Beijing 100191}
\affiliation{University of Bonn, 53115 Bonn}
\affiliation{Budker Institute of Nuclear Physics SB RAS and Novosibirsk State University, Novosibirsk 630090}
\affiliation{Faculty of Mathematics and Physics, Charles University, 121 16 Prague}
%%%\affiliation{Chiba University, Chiba 263-8522}
%%%\affiliation{Chonnam National University, Kwangju 660-701}
\affiliation{University of Cincinnati, Cincinnati, Ohio 45221}
\affiliation{Deutsches Elektronen--Synchrotron, 22607 Hamburg}
%%%\affiliation{University of Florida, Gainesville, Florida 32611}
\affiliation{Department of Physics, Fu Jen Catholic University, Taipei 24205}
\affiliation{Justus-Liebig-Universit\"at Gie\ss{}en, 35392 Gie\ss{}en}
\affiliation{Gifu University, Gifu 501-1193}
%%%\affiliation{II. Physikalisches Institut, Georg-August-Universit\"at G\"ottingen, 37073 G\"ottingen}
\affiliation{SOKENDAI (The Graduate University for Advanced Studies), Hayama 240-0193}
%%%\affiliation{Gyeongsang National University, Chinju 660-701}
\affiliation{Hanyang University, Seoul 133-791}
\affiliation{University of Hawaii, Honolulu, Hawaii 96822}
\affiliation{High Energy Accelerator Research Organization (KEK), Tsukuba 305-0801}
%%%\affiliation{Hiroshima Institute of Technology, Hiroshima 731-5193}
\affiliation{IKERBASQUE, Basque Foundation for Science, 48013 Bilbao}
%%%\affiliation{University of Illinois at Urbana-Champaign, Urbana, Illinois 61801}
%%%\affiliation{Indian Institute of Technology Bhubaneswar, Satya Nagar 751007}
%%%\affiliation{Indian Institute of Technology Guwahati, Assam 781039}
\affiliation{Indian Institute of Technology Madras, Chennai 600036}
\affiliation{Indiana University, Bloomington, Indiana 47408}
\affiliation{Institute of High Energy Physics, Chinese Academy of Sciences, Beijing 100049}
\affiliation{Institute of High Energy Physics, Vienna 1050}
\affiliation{Institute for High Energy Physics, Protvino 142281}
%%%\affiliation{Institute of Mathematical Sciences, Chennai 600113}
\affiliation{INFN - Sezione di Torino, 10125 Torino}
\affiliation{Institute for Theoretical and Experimental Physics, Moscow 117218}
\affiliation{J. Stefan Institute, 1000 Ljubljana}
\affiliation{Kanagawa University, Yokohama 221-8686}
\affiliation{Institut f\"ur Experimentelle Kernphysik, Karlsruher Institut f\"ur Technologie, 76131 Karlsruhe}
%%%\affiliation{Kavli Institute for the Physics and Mathematics of the Universe (WPI), University of Tokyo, Kashiwa 277-8583}
%%%\affiliation{Kennesaw State University, Kennesaw GA 30144}
\affiliation{King Abdulaziz City for Science and Technology, Riyadh 11442}
\affiliation{Department of Physics, Faculty of Science, King Abdulaziz University, Jeddah 21589}
\affiliation{Korea Institute of Science and Technology Information, Daejeon 305-806}
\affiliation{Korea University, Seoul 136-713}
%%%\affiliation{Kyoto University, Kyoto 606-8502}
\affiliation{Kyungpook National University, Daegu 702-701}
\affiliation{\'Ecole Polytechnique F\'ed\'erale de Lausanne (EPFL), Lausanne 1015}
\affiliation{Faculty of Mathematics and Physics, University of Ljubljana, 1000 Ljubljana}
\affiliation{Luther College, Decorah, Iowa 52101}
\affiliation{University of Maribor, 2000 Maribor}
\affiliation{Max-Planck-Institut f\"ur Physik, 80805 M\"unchen}
\affiliation{School of Physics, University of Melbourne, Victoria 3010}
\affiliation{Moscow Physical Engineering Institute, Moscow 115409}
\affiliation{Moscow Institute of Physics and Technology, Moscow Region 141700}
\affiliation{Graduate School of Science, Nagoya University, Nagoya 464-8602}
\affiliation{Kobayashi-Maskawa Institute, Nagoya University, Nagoya 464-8602}
%%%\affiliation{Nara University of Education, Nara 630-8528}
\affiliation{Nara Women's University, Nara 630-8506}
\affiliation{National Central University, Chung-li 32054}
\affiliation{National United University, Miao Li 36003}
\affiliation{Department of Physics, National Taiwan University, Taipei 10617}
\affiliation{H. Niewodniczanski Institute of Nuclear Physics, Krakow 31-342}
%%%\affiliation{Nippon Dental University, Niigata 951-8580}
\affiliation{Niigata University, Niigata 950-2181}
\affiliation{University of Nova Gorica, 5000 Nova Gorica}
\affiliation{Osaka City University, Osaka 558-8585}
%%%\affiliation{Osaka University, Osaka 565-0871}
\affiliation{Pacific Northwest National Laboratory, Richland, Washington 99352}
%%%\affiliation{Panjab University, Chandigarh 160014}
\affiliation{Peking University, Beijing 100871}
\affiliation{University of Pittsburgh, Pittsburgh, Pennsylvania 15260}
%%%\affiliation{Punjab Agricultural University, Ludhiana 141004}
%%%\affiliation{Research Center for Electron Photon Science, Tohoku University, Sendai 980-8578}
%%%\affiliation{Research Center for Nuclear Physics, Osaka University, Osaka 567-0047}
%%%\affiliation{RIKEN BNL Research Center, Upton, New York 11973}
%%%\affiliation{Saga University, Saga 840-8502}
\affiliation{University of Science and Technology of China, Hefei 230026}
\affiliation{Seoul National University, Seoul 151-742}
%%%\affiliation{Shinshu University, Nagano 390-8621}
\affiliation{Soongsil University, Seoul 156-743}
\affiliation{University of South Carolina, Columbia, South Carolina 29208}
\affiliation{Sungkyunkwan University, Suwon 440-746}
\affiliation{School of Physics, University of Sydney, NSW 2006}
\affiliation{Department of Physics, Faculty of Science, University of Tabuk, Tabuk 71451}
\affiliation{Tata Institute of Fundamental Research, Mumbai 400005}
\affiliation{Excellence Cluster Universe, Technische Universit\"at M\"unchen, 85748 Garching}
%%%\affiliation{Toho University, Funabashi 274-8510}
%%%\affiliation{Tohoku Gakuin University, Tagajo 985-8537}
\affiliation{Tohoku University, Sendai 980-8578}
\affiliation{Department of Physics, University of Tokyo, Tokyo 113-0033}
\affiliation{Tokyo Institute of Technology, Tokyo 152-8550}
\affiliation{Tokyo Metropolitan University, Tokyo 192-0397}
%%%\affiliation{Tokyo University of Agriculture and Technology, Tokyo 184-8588}
\affiliation{University of Torino, 10124 Torino}
%%%\affiliation{Toyama National College of Maritime Technology, Toyama 933-0293}
%%%\affiliation{Utkal University, Bhubaneswar 751004}
\affiliation{CNP, Virginia Polytechnic Institute and State University, Blacksburg, Virginia 24061}
\affiliation{Wayne State University, Detroit, Michigan 48202}
\affiliation{Yamagata University, Yamagata 990-8560}
\affiliation{Yonsei University, Seoul 120-749}
   \author{C.~Oswald}\affiliation{University of Bonn, 53115 Bonn} % Bonn
   \author{P.~Urquijo}\affiliation{School of Physics, University of Melbourne, Victoria 3010} % Melbourne
   \author{J.~Dingfelder}\affiliation{University of Bonn, 53115 Bonn} % Bonn
  \author{A.~Abdesselam}\affiliation{Department of Physics, Faculty of Science, University of Tabuk, Tabuk 71451} % Tabuk
  \author{I.~Adachi}\affiliation{High Energy Accelerator Research Organization (KEK), Tsukuba 305-0801}\affiliation{SOKENDAI (The Graduate University for Advanced Studies), Hayama 240-0193} % KEK
% \author{K.~Adamczyk}\affiliation{H. Niewodniczanski Institute of Nuclear Physics, Krakow 31-342} % Krakow
  \author{H.~Aihara}\affiliation{Department of Physics, University of Tokyo, Tokyo 113-0033} % Tokyo
  \author{S.~Al~Said}\affiliation{Department of Physics, Faculty of Science, University of Tabuk, Tabuk 71451}\affiliation{Department of Physics, Faculty of Science, King Abdulaziz University, Jeddah 21589} % Tabuk
% \author{K.~Arinstein}\affiliation{Budker Institute of Nuclear Physics SB RAS and Novosibirsk State University, Novosibirsk 630090} % BINP
% \author{Y.~Arita}\affiliation{Graduate School of Science, Nagoya University, Nagoya 464-8602} % Nagoya
  \author{D.~M.~Asner}\affiliation{Pacific Northwest National Laboratory, Richland, Washington 99352} % PNNL
% \author{T.~Aso}\affiliation{Toyama National College of Maritime Technology, Toyama 933-0293} % Toyama
% \author{V.~Aulchenko}\affiliation{Budker Institute of Nuclear Physics SB RAS and Novosibirsk State University, Novosibirsk 630090} % BINP
  \author{T.~Aushev}\affiliation{Moscow Institute of Physics and Technology, Moscow Region 141700}\affiliation{Institute for Theoretical and Experimental Physics, Moscow 117218} % ITEP
  \author{R.~Ayad}\affiliation{Department of Physics, Faculty of Science, University of Tabuk, Tabuk 71451} % Tabuk
% \author{T.~Aziz}\affiliation{Tata Institute of Fundamental Research, Mumbai 400005} % Tata
% ...
% Ljubljana
% \author{N.~Zwahlen}\affiliation{\'Ecole Polytechnique F\'ed\'erale de Lausanne (EPFL), Lausanne 1015} % Lausanne
% \author{O.~Zyukova}\affiliation{Budker Institute of Nuclear Physics SB RAS and Novosibirsk State University, Novosibirsk 630090} % BINP
\collaboration{The Belle Collaboration}

\noaffiliation

\maketitle

{\renewcommand{\thefootnote}{\fnsymbol{footnote}}}
\setcounter{footnote}{0}

\section{Introduction}

Analyses of semileptonic decays $B \to X_c \ell \nu$, where $X_c$ denotes a hadronic final state with a charm quark, play an important role in the determination of the CKM matrix element $|V_{cb}|$. The extraction of $|V_{cb}|$ from the measured decay rates relies on form factors that describe the accompanying strong interaction processes. Measurements of semileptonic $B_s$ decays provide complementary information to test and validate the QCD calculations of these form factors. Since large $B_s$ samples have become available at Belle and the experiments at the Large Hadron Collider, the interest in the topic of semileptonic $B_s$ decays has intensified recently. Theoretical predictions of form factors and branching fractions are based on QCD sum rules~\cite{Li:2009wq,Azizi:2008tt,Azizi:2008vt,Blasi:1993fi}, lattice QCD~\cite{Atoui:2013zza,Bailey:2012rr} and constituent quark models~\cite{Fan:2013kqa, Faustov:2012mt,Li:2010bb,Zhang:2010ur,Zhao:2006at,
Chen:2011ut,Segovia:2011dg,Albertus:2014bfa,Bhol:2014jta}.
The predicted exclusive branching fractions vary from 1.0\% to 3.2\%  for $B_s \to D_s \ell \nu$ decays and from 4.3\% to 7.6\%  for 
$B_s \to D_s^* \ell \nu$ decays. There are also predictions for the modes with higher excitations of the $D_s$ meson, denoted hereinafter by ``$D_s^{**}$''. The LHCb and D\O~experiments have measured the semi-inclusive branching fractions of the decays $B_s \to D_{s1}(2536)X\mu^+\nu$ and $B_s \to D_{s2}^*(2573)X\mu^+\nu$, where the $D_s^{**}$ mesons were reconstructed in $D^{(*)}K$ final states~\cite{Abazov:2007wg,Aaij:2011ju}. The inclusive semileptonic branching fraction of $B_s \to X \ell \nu$ decays was recently measured by Belle and BaBar~\cite{Oswald:2012yx,Lees:2011ji} and found to be in agreement with the expectations from SU(3) flavor symmetry~\cite{bigi2011,Gronau2010}. We report here the first measurements of the semi-inclusive branching fractions $\mathcal{B}(B_s \to D_s X \ell \nu)$ and $\mathcal{B}(B_s \to D_s^* X \ell \nu)$ using the Belle $\Upsilon(5S)$ dataset. The number of $B_s^{(*)}\bar{B}_s^{(*)}$ pairs in the dataset, 
\begin{equation}
N_{B_s\bar{B}_s} = \sigma(e^+e^- \to B_s^{(*)}\bar{B}_s^{(*)}) \cdot \mathcal{L}_{\Upsilon(5S)}\,,
\end{equation}
where $\sigma(e^+e^- \to B_s^{(*)}\bar{B}_s^{(*)})$ is the production cross-section and $\mathcal{L}_{\Upsilon(5S)}$ is the integrated luminosity, is the limiting systematic uncertainty in this measurement and other untagged $B_s$ measurements at Belle~\cite{Oswald:2013tna}. The value $N_{B_s\bar{B}_s} = (7.1 \pm 1.3)~\times~10^6$ was obtained from a measurement of the inclusive $D_s$ yield in the dataset~\cite{NBs}. The measured $B_s \to D_s X \ell \nu$ yield, together with an estimate for the branching fraction $\mathcal{B}(B_s \to D_s X \ell \nu)$, provides an alternative way to determine $N_{B_s\bar{B}_s}$. A similar approach was already pursued by the LEP experiments~\cite{Buskulic:1995bd,Abreu:1992rv,Acton:1992zq} and LHCb~\cite{Aaij:2011jp}. 

\section{Detector, data sample and simulation}

The Belle detector located at the KEKB asymmetric-energy $e^+e^-$ collider~\cite{KEKB} is a large-solid-angle magnetic spectrometer that consists of a silicon vertex detector, a 50-layer central drift chamber (CDC), an array of aerogel threshold Cherenkov counters (ACC),  a barrel-like arrangement of time-of-flight scintillation counters (TOF), and an electromagnetic calorimeter (ECL) comprised of CsI(Tl) crystals located inside a superconducting solenoid coil that provides a 1.5~T magnetic field.  An iron flux-return located outside of the coil is instrumented to detect $K_L^0$ mesons and to identify muons (KLM). The detector is described in detail elsewhere~\cite{Belle}.

This analysis uses a dataset with an integrated luminosity of $\mathcal{L}_{\Upsilon(5S)}=(121.4 \pm 0.8)~{\rm fb^{-1}}$ collected at a center-of-mass (CM) energy of $\sqrt{s} = \unit[10.86]{GeV}$~\cite{speedoflight}, corresponding to the mass of the $\Upsilon(5S)$ resonance. The $B_s$ mesons are produced in pairs in the following production modes, with the respective production fractions given in parentheses: $B_s^* \bar{B}_s^*$ ($(87.8 \pm 1.5)\%$), $B_s^* \bar{B}_s$ ($(6.7 \pm 1.2)\%$) and $B_s \bar{B}_s$ ($(2.6 \pm 2.6)\%$)~\cite{PDBook}. All production modes are considered for the analysis. Moreover, we use a $62.8~{\rm fb^{-1}}$ sample collected below the production threshold for open $B$ production to study the continuum processes $e^+e^- \to q\bar{q}$ ($q = u,d,s,c$).

A sample of simulated events with a size corresponding to six times the integrated data luminosity is generated using Monte Carlo (MC) techniques. The simulated data emulate the different types of events produced at the $\Upsilon(5S)$ CM energy, comprising events with $B$ and $B_s$ decays, bottomonium production and the $q\bar{q}$ continuum processes. The events are generated  with the \texttt{EvtGen} package~\cite{evtgen} and are processed through a \texttt{GEANT}~\cite{geant3} based detector simulation. Final state photon radiation is added with the \texttt{PHOTOS} package~\cite{photos}. 

The branching fractions in the simulation are set to the latest averages from the Particle Data Group~\cite{PDBook}. However, for semileptonic $B_s$ decays only measurements of the $D_{s1}(2536)$ and $D_{s2}^*(2573)$ modes are available, so we use instead the calculations from Faustov and Galkin~\cite{Faustov:2012mt}, who predict the full set of branching fractions and thus provide a self-consistent picture of the semileptonic width. The $B_s$ semileptonic decay modes considered in this analysis, with their corresponding branching fractions given in parentheses, are: $D_s$ (2.1\%), $D_s^*$ (5.3\%),  $D_{s1}(2536)$ (0.84\%), $D_{s0}^*(2317)$ (0.36\%), $D_{s1}(2460)$ (0.19\%) and $D_{s2}^*(2573)$ (0.67\%). The decays $B_{s}^0 \to D_{s}^{(*)}\ell\nu$ are described by the Caprini-Lellouch-Neubert model~\cite{Caprini:1997mu}, based on heavy quark effective theory~\cite{Neubert:1993mb}. Assuming SU(3) flavor symmetry, the form factors of the semileptonic $B_s$ decays are taken to be identical to the ones measured in the corresponding $B$ decays~\cite{hfag}; we use the following values of the form factor parameters: $\rho_D = 1.186$ for $B_s \to D_s \ell \nu$ decays, and $\rho = 1.207$, $R_1 = 1.403$, $R_2 = 0.854$ for $B_s \to D_s^* \ell \nu$ decays. The $B_s^0 \to D_s^{**} \ell \nu$ decays are described by the Leibovich-Ligeti-Stewart-Wise (LLSW) model~\cite{llsw} originally developed for $B \to D^{**} \ell \nu$ decays. We replace in this model the $B$ and $D^{**}$ masses by the $B_s$ and $D_s^{**}$ masses, respectively. The nominal branching fractions for the $D_s^{**}$ decays in this analysis are listed in Table~\ref{tab:dsststbfs}.

\begin{table}
\caption{Nominal branching fractions of $D_s^{**}$ decays to different final states in the MC simulation. The branching fraction of the $D_s^* \to D_s X$ decays is set to 100\% and this crossfeed is included in the calculation of the branching fractions to the $D_s X$ final state.}
\begin{tabular}{lrrr}
\hline \hline
& \multicolumn{3}{c}{Branching fraction [\%]} \\
$Y$ & $Y \to D_s X$ & \hspace{0.3cm} $Y \to D_s^* X$ & \hspace{0.3cm} $Y \to D^{(*)} K$ \\
\hline
$D_{s0}^*(2317)$ & 100 & 63 & 0 \\
$D_{s1}(2460)$ & 100 & 3 & 0 \\
$D_{s1}(2536)$ & 0 & 0 & 100 \\
$D_{s2}^*(2573)$ & 0 & 0 & 100 \\
\hline \hline
\end{tabular}
\label{tab:dsststbfs}
\end{table}

\section{Analysis overview}

This analysis is based on samples of reconstructed $D_s^- \ell^+$ and $D_s^{*-} \ell^+$ pairs~\cite{CC}. Incorrectly reconstructed $D_s$ and $D_s^*$ candidates constitute a large background in the analysis. We therefore perform fits to the $D_s^{(*)}$ mass distributions to determine the yields of events with correctly reconstructed $D_s^{(*)}$ mesons. These events contain the following signal and background categories:
\begin{enumerate}
\item $e^+e^- \to c\bar{c}$ continuum; 
\item $B \to D_s^{(*)} K \ell \nu$ decays, which have a  branching fraction of $(6.1 \pm  1.0)\times10^{-4}$~\cite{PDBook,Stypula:2012mf,delAmoSanchez:2010pa};
\item \emph{opposite-$B_{(s)}$} background, where a lepton candidate is combined with a $D_s^{(*)}$ meson from the second $B_s$ in the event; the lepton candidate can be either a primary lepton from a $B_{(s)} \to X \ell \nu$ decay, a lepton originating from a secondary decay or a misidentified hadron track; 
\item \emph{same-$B_{(s)}$} background from secondary leptons and from hadron tracks misidentified as leptons, which stem from the decay of the same $B_s$ meson as the reconstructed $D_s^{(*)}$ meson; 
\item \emph{signal}: in the $D_s X \ell \nu$ channel, the signal comprises $B_s \to D_s \ell \nu$ decays and  crossfeed from $B_s \to D_s^* \ell \nu$ and $B_s \to D_s^{**} \ell \nu$ decays; in the $D_s^* X \ell\nu$ channel, the dominant signal contributions are $B_s \to D_s^* \ell \nu$ decays with a small crossfeed contribution from $B_s \to D_s^{**} \ell \nu$ decays. 
\end{enumerate}
The continuum background is estimated using off-resonance data, and the $B \to D_s^{(*)} K \ell \nu$ background is estimated from MC simulation. We use the kinematic properties of the reconstructed decay to determine the normalisations of the other three components from data. 
For this, we consider the lepton momentum in the CM system of the $e^+e^-$ collision, $p^*_\ell$, and the variable
\begin{equation}
X_\text{mis} = \frac{E^*_{B_s} - (E^*_{D_s \ell} + p^*_{D_s \ell})}{p^*_{B_s}}\,,
\label{eq:xmis}
\end{equation}
where $E^*_{B_s}$ is the energy of the $B_s$ meson in the CM system approximated by $\sqrt{s}/2$; $p^*_{B_s}$ is the momentum of the $B_s$ meson in the CM system approximated by $\sqrt{s/4 - m_{B_s}^2}$; $E^*_{D_s \ell} = E^*_\ell + E^*_{D_s}$ is the sum of the reconstructed energies in the CM system and $p^*_{D_s \ell} = |\vec{p}^*_\ell + \vec{p}^*_{D_s}|$ is the absolute value of the sum of the reconstructed lepton and $D_s$ momenta in the CM system. When the $D_s$ meson and the lepton candidate stem from the decay of the same $B_s$ meson, $X_\text{mis}$ takes values larger than $-1$ because the momentum of the unreconstructed $B_s$ decay products, $p^*_\text{other}$, is constrained by the triangle inequality $p^*_{B_s} - p^*_{D_s\ell} \leq p^*_\text{other}$ and by $p^*_\text{other} \leq E^*_\text{other} = E^*_{B_s} - E^*_{D_s\ell}$. We divide the data samples into three regions:
\begin{description}
\item[A:] $X_\text{mis} < -1$,
\item[B:] $X_\text{mis} \geq -1$ and $p^*_\ell < 1.4~{\rm GeV}$,
\item[C:] $X_\text{mis} \geq -1$ and $p^*_\ell \geq 1.4~{\rm GeV}$.
\end{description}
As these regions are later used to determine the signal yields, we refer to them as ``counting regions'' in the following. Region A contains only \emph{opposite-$B_{(s)}$} background and can be used to determine the normalisation of this background. The normalisation of the other two components can be extracted from the measured yields in regions B and C, which have an enhanced fraction of \emph{same-$B_{(s)}$} background and \emph{signal} events, respectively. The boundary $p^*_\ell = 1.4~{\rm GeV}$ is chosen to achieve approximately equal event yields in regions B and C. The analysis is insensitive to the modeling of the $X_\text{mis}$ distribution for signal decays, which depends on the mass of the $B_s^*$ meson, $m_{B_s^*}$, and is thus strongly influenced by the poor precision on $m_{B_s^*}$. The semi-inclusive branching fractions are obtained from the relation
\begin{equation}
\mathcal{B}(B_s \to D_s^{(*)} X \ell \nu) = \frac{N_\text{sig}}{2 \cdot N_{B_s\bar{B}_s} \epsilon \mathcal{B}_{D_s^{(*)}}}\,,
\label{eq:brafradef}
\end{equation} 
where $N_\text{sig}$ is the measured signal yield, $\epsilon$ is the average signal efficiency and $\mathcal{B}_{D_s^{(*)}}$ is the branching fraction of the reconstructed $D_s^{(*)}$ decay mode:

\begin{equation}
\mathcal{B}_{D_s} = \mathcal{B}(D_s^- \to \phi \pi^-;~ \phi \to K^+K^-)\,,
\end{equation}
\begin{equation}
\mathcal{B}_{D_s^*} = \mathcal{B}(D_s^{*-} \to D_s^- \gamma) \cdot \mathcal{B}_{D_s} \,.
\end{equation}

\section{Event selection}

We select tracks originating from the interaction region by requiring $|dz| < 2.0\,{\rm cm}$ and $dr < 0.5\,{\rm cm}$, where $dz$ and $dr$ are the impact parameters along the $e^+$ beam and in the transverse plane, respectively. Kaon or pion hypotheses are assigned to the tracks based on a likelihood combining the information from the Cherenkov light yield in the ACC, the time-of-flight information of the TOF and the specific ionization ${\rm d}E/{\rm d} x$ in the CDC. The kaon (pion) identification efficiency for tracks with a typical momentum of $0.75\,{\rm GeV}$ is about 96\% (92\%), while the rate of pions (kaons) being misidentified as kaons (pions) is 7\% (2\%). The kaon and pion candidates are used to reconstruct $D_s$ mesons in the high-purity decay channel $D_s^- \to \phi \pi^-;~ \phi \to K^+K^-$. A $D_s$ candidate is retained in the analysis if it has a reconstructed mass, $M_{KK\pi}$, within a $\unit[\pm 65]{MeV}$ window around the nominal $D_s$ mass, $m_{D_s} = \unit[1968.5]{MeV}$~\cite{PDBook}, that includes large enough sidebands to determine the combinatorial background of random $KK\pi$ combinations. The reconstructed di-kaon invariant mass, $M_{KK}$, is required to be in the mass window between $1004$ and $1034\,{\rm MeV}$, corresponding to three times the FWHM of the reconstructed $\phi$ mass peak. To suppress combinatorial background, we impose the criterion $|\cos \theta_\text{hel}| > 0.3$ on the helicity angle, defined as the angle between the momentum of the $D_s$ and the $K^-$ in the rest frame of the $\phi$ resonance.

The $D_s$ candidates with a reconstructed mass, $M_{KK\pi}$, within the range between $1953.5$ and $1983.5\,{\rm MeV}$, corresponding to three times the RMS of the $D_s$ mass peak, are utilized for the reconstruction of $D_s^*$ candidates in the dominant decay channel $D_s^* \to D_s \gamma$. Photon candidates are reconstructed from ECL clusters that are not attributed to a track candidate. The photon candidate must have a minimum energy of $\unit[125]{MeV}$ in the lab frame and the ratio of the energy deposit in the central $3\times3$ cells of the ECL cluster to the energy deposit in the central $5\times5$ cells must be at least $90\%$. To veto photons from $\pi^0$ decays, we combine the photon candidate with any other photon candidate in the detector and require that the invariant mass of the two photons differs from the nominal $\pi^0$ mass~\cite{PDBook} by more than $\unit[5]{MeV}$, corresponding to about 0.8 times the experimental resolution. The angle between the $D_s$ meson and the photon in the lab frame is typically less than $90^{\circ}$, so only candidates fulfilling this requirement are retained. The $D_s^*$ candidates whose mass difference between the reconstructed $D_s^*$ and $D_s$ candidates, $\Delta M = M_{KK\pi\gamma} - M_{KK\pi}$, lies between $78.8$ and $208.8\,{\rm MeV}$ are retained.

Electron and muon candidates are reconstructed from tracks that are not used for the $D_s^{(*)}$ reconstruction. Electrons are selected based on the position matching between the track and the ECL cluster, the ratio of the energy measured in the ECL to the charged track momentum, the transverse ECL shower shape, specific ionization in the CDC and the ACC light yield. Muons are identified using their penetration depth and the transverse scattering in the KLM. Hadron tracks misidentified as leptons and leptons from secondary decays tend to have lower momenta than primary leptons and are suppressed by rejecting lepton candidates with a momentum in the lab frame below $\unit[900]{MeV}$. The electron (muon) identification efficiency in the selected momentum region is better than 89\% (82\%) and the probability that a charged pion or kaon track is misidentified as an electron (muon) is below 1\% (2\%). Leptons, $\ell^+$, from $J/\psi \to \ell^+\ell^-$ decays are vetoed by requiring $|M_{\ell^+ h^-} - m_{J/\psi}| < \unit[5]{MeV}$, where $M_{\ell^+ h^-}$ is the invariant mass of the lepton and any accepted track of the opposite charge, $h^-$, to which we assign the $\ell$ mass hypothesis. Furthermore, electrons are rejected if they are likely to stem from photon conversions, $|M_{\ell^+ h^-}| < \unit[100]{MeV}$, or Dalitz $\pi^0$ decays, $|M_{\ell^+ h^- \gamma} - m_{\pi^0}| < \unit[32]{MeV}$.

We form a signal $B_s^0$ candidate by pairing a $D_s^{(*)-}$ candidate with an oppositely charged lepton candidate $\ell^+$. To suppress background from $c\bar{c}$ continuum, we reject events where the normalised $D_s$ momentum, $x(D_s) = p^*(D_s) / \sqrt{s/4 - m_{D_s}^2}$, is larger than $0.5$ (for explanations, see Ref.~\cite{Oswald:2012yx}). Further suppression of the $c\bar{c}$ continuum background is achieved by rejecting events with a jet-like topology characterised by $|\cos \theta_\text{thrust}| > 0.8$, where $\theta_\text{thrust}$ is the thrust angle defined by the two thrust axes maximizing the projection of the momenta of the tracks and photon candidates of the $B_s^0$ candidate and the rest of the event, respectively.

After applying the selection criteria, $7.9\%$ ($0.4\%$) of the events contain more than one (two) $D_s^+\ell^-$ candidate(s). We perform a $\chi^2$ fit to the vertex of the three tracks used for $D_s$ reconstruction and select the candidate with the best goodness-of-fit. This approach selects a correct candidate in 80\% of the cases. The selected $D_s$ candidate in an event is used for $D_s^*$ reconstruction; in $36.2\%$ ($9.7\%$) of the events, more than one (two) $D_s \gamma$ combinations meet the $D_s^*$ requirements. We choose the photon candidate with the highest energy fraction deposited in the central $3\times3$ cells of a $5\times5$ cell ECL cluster. In the case that more than one photon candidate deposits all of its energy in the central $3\times3$ cells of the cluster, the candidate with the higher energy in the lab frame is selected. If two or more lepton candidates pass all of these selection criteria ($2.1 \%$ of all events), we choose a random lepton candidate.

\section{Fit results}

\subsection{{\boldmath $D_s$} fits}

\begin{figure*}
\centering
\subfigure[$D_s^-e^+$]{
\includegraphics[width=0.9\textwidth]{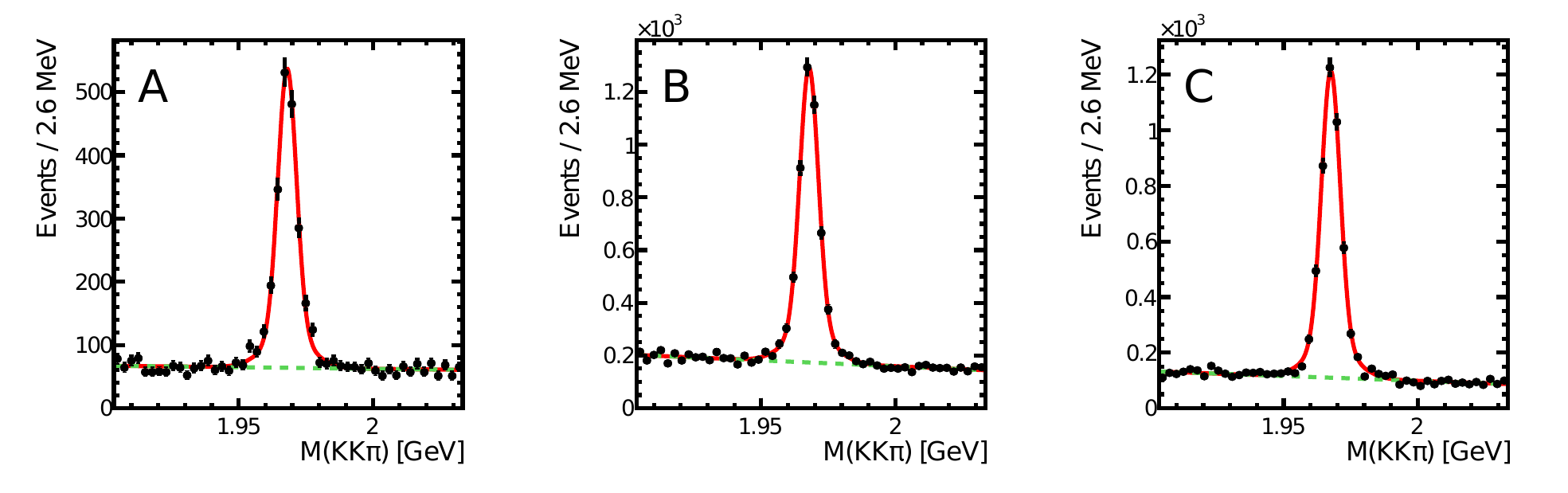}}
\subfigure[$D_s^-\mu^+$]{
\includegraphics[width=0.9\textwidth]{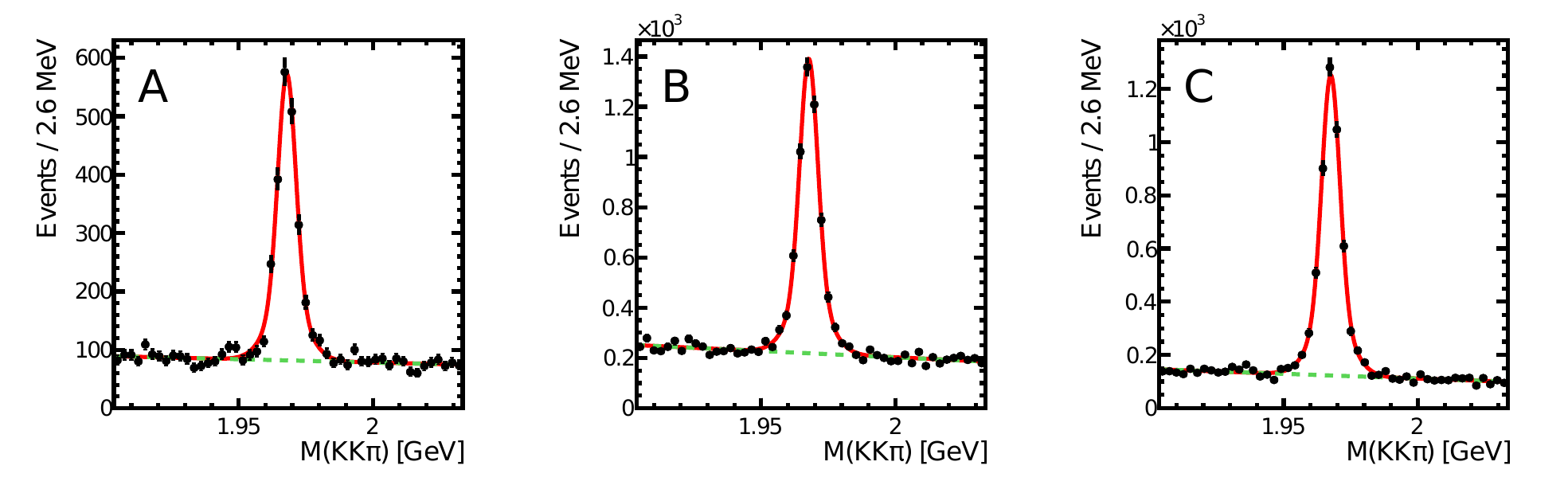}}
\subfigure[$D_s^{*-}e^+$]{
\includegraphics[width=0.9\textwidth]{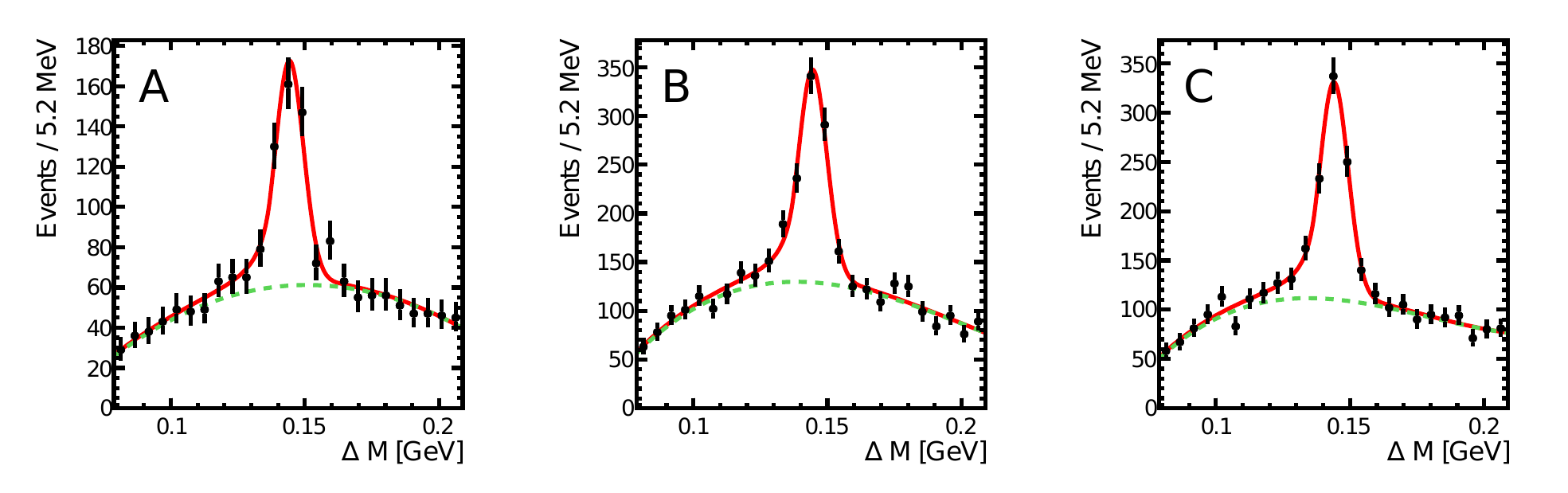}}
\subfigure[$D_s^{*-}\mu^+$]{
\includegraphics[width=0.9\textwidth]{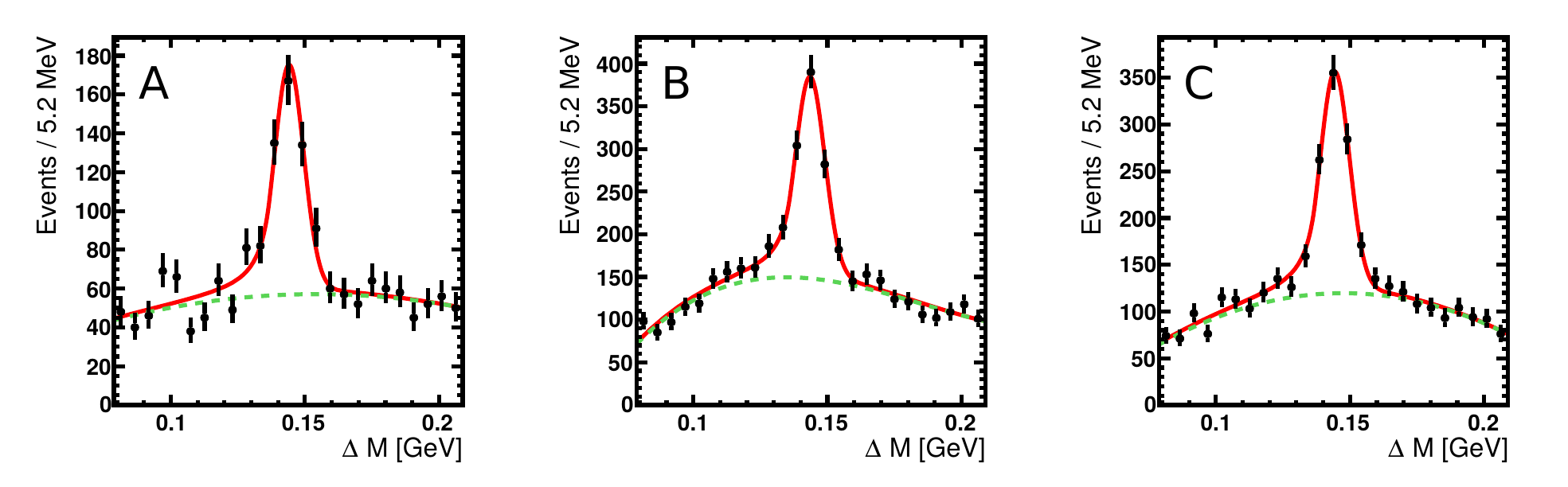}}
\caption{The $M_{KK\pi}$ distributions for $D_s^{-} \ell^+$ events and $\Delta M$ distributions for $D_s^{*-} \ell^+$ events reconstructed in the
$\Upsilon(5S)$ data for the three counting regions. The black points with uncertainty bars are the data, the red solid curve represents the total fit result, and the green dashed line is the fitted background component.}
\label{fig:mass_fits}
\end{figure*}

We determine the yields of correctly reconstructed $D_s$ mesons with binned extended maximum likelihood fits to the reconstructed $D_s$ mass, $M=M_{KK\pi}$, in 50 equal bins, indexed by $j$. The probability density function (PDF) of correctly reconstructed $D_s$ mesons, $\mathcal{P}_\text{sig}(M)$, is modeled by the sum of two Gaussian functions with a common mean. The PDF of the combinatorial background, $\mathcal{P}_\text{bkg}(M)$, is a first-order Chebychev polynomial. We do not determine the shape parameters from simulation, but rather allow them to vary as free parameters in the fit. The $D_s$ mass fits are performed simultaneously in the three counting regions ($i = A, B, C$) defined above. The width of the first Gaussian function, $\sigma_1$, the ratio of the widths of the two Gaussian functions, $r_\sigma$, and the ratio of the normalisations of the two Gaussian functions, $r_N$, are common fit parameters in all three regions. The means of the Gaussian functions, $\mu_i$, and the slopes of the polynomials describing the background, $b_i$, are fitted in each counting region individually. The likelihood function is:
\begin{equation}
L(\nu^\text{sig}, \nu^\text{bkg}, \theta) =
\prod_{i=A,B,C} 
\frac{\nu_i^{n_i}}{n_i!}{\rm e}^{-\nu_i}
\prod_{\text{bins j}}
\frac{\nu_{ij}^{n_{ij}}}{n_{ij}!}{\rm e}^{-\nu_{ij}}\,,
\label{eq:likelihood}
\end{equation}
where $\nu^\text{sig} =  (\nu_A^\text{sig}, \nu_B^\text{sig}, \nu_C^\text{sig})$ and $\nu^\text{bkg} =  (\nu_A^\text{bkg}, \nu_B^\text{bkg}, \nu_C^\text{bkg})$ is the vector of signal and background yields in the three counting regions, $\theta = (\sigma_1, r_\sigma, r_N, \mu_A, \mu_B, \mu_C, b_A, b_B, b_C)$ are the shape parameters for the signal and background PDFs, and $n_{ij}$ and $\nu_{ij}$ are the observed and expected event yields in bin $j$ of counting region $i$, respectively, with $n_i = \sum_j n_{ij}$ and $\nu_i = \sum_j \nu_{ij}$. The expected event yield, $\nu_{ij}$, is a function of $\nu_i^\text{sig}$,$\nu_i^\text{bkg}$ and $\theta$:
\begin{equation}
\nu_{ij} = \int\limits_{M_{j,\text{min}}}^{M_{j,\text{max}}} \left[ \nu_i^\text{sig} \mathcal{P}_\text{sig} (M) +
\nu_i^\text{bkg} \mathcal{P}_\text{bkg} (M) \right]
{\rm d}M \,.
\label{eq:expyield}
\end{equation}
Figures~\ref{fig:mass_fits}~(a) and (b) show the $KK\pi$ mass distributions together with the fit results.

\subsection{{\boldmath $D_s^*$} fits}

The $D_s^*$ yields are determined from binned extended maximum likelihood fits to the mass difference $\Delta M$  in 25 equal bins, indexed by $j$. The combinatorial background is modeled by a third-order Chebyshev polynomial, $\mathcal{P}_\text{bkg}(\Delta M)$, whose parameters are constrained to the values obtained from fits to simulated background distributions. Since the background shapes vary for the different counting regions, the shape parameters are determined for each counting region separately. The signal peak is modeled by the sum of a Gaussian function and a Crystal Ball function~\cite{CBall} to account for energy loss due to material in front of the calorimeter:
\begin{equation*}
\begin{array}{ll}
\mathcal{P}_\text{sig}(\Delta M) \propto &
r_N \exp\left(- \frac{(\Delta M - \mu)^2}{(r_\sigma \cdot \sigma)^2}\right) \\
& + \begin{cases}
\exp \left(-\frac{(\Delta M  - \mu)^2}{2 \sigma^2}\right) & \text{if}~~\frac{\Delta M  - \mu}{\sigma} > - \alpha 
\\
\frac{
\left(\frac{\beta}{\alpha}\right)^\beta \cdot {\rm e}^{-\alpha^2/2}}{
\left(\frac{\beta}{\alpha} - \alpha - \frac{\Delta M  - \mu}{ \sigma} \right)^{\beta}}
& \text{if}~~\frac{\Delta M  - \mu}{ \sigma} \leq - \alpha\,.
\end{cases}
\end{array}
\end{equation*}
A common mean, $\mu$, is used for both the Gaussian and the Crystal Ball functions. We perform a fit to the simulated signal distribution and fix the parameters $r_N$, $r_\sigma$, $\alpha$ and $n$ at the obtained values. The width $\sigma$ and the mean of the signal peak $\mu$ are  varied in the fit to data; the parameter $\sigma$ is fitted simultaneously in all counting regions while $\mu$ is fitted individually for each counting region. The likelihood function is constructed analogous to Eqs.~(\ref{eq:likelihood}) and~(\ref{eq:expyield}) with additional factors, to implement the constraints of the background PDF parameters taking into account their correlations. The results of the $\Delta M$ fits in the different counting regions are presented in Figs.~\ref{fig:mass_fits}~(c) and (d).

\subsection{Background subtraction}

To estimate the $c\bar{c}$ continuum background, the $D_s$ and $D_s^*$ yields are measured in $D_s^-\ell^+$ and $D_s^{*-} \ell^+$ samples reconstructed in the off-resonance data. Since the size of the off-resonance data sample is not sufficient to determine the shape parameters in the fits, they are fixed to the values obtained in the fits to $\Upsilon(5S)$ data in the corresponding counting region. The CM energy, $\sqrt{s}$, in the expression for the $X_\text{mis}$ variable in Eq.~(\ref{eq:xmis}) is replaced by a constant value of $10.876~{\rm GeV}$ because, otherwise, the denominator would not be defined. The $c\bar{c}$ continuum yields from the fits to off-resonance data are multiplied by the scale factor $S = (\mathcal{L}_{\Upsilon(5S)}/s_{\Upsilon(5S)}) / (\mathcal{L}_{\text{off}}/s_{\text{off}}) = 1.81 \pm 0.02$ to account for the differences in integrated luminosities, $\mathcal{L}$, and the $1/s$ dependence of the $e^+e^- \to c\bar{c}$ cross section. Additionally, a shape correction for differences of the yields in the counting regions between off-resonance and $\Upsilon(5S)$ data is determined from MC simulation and applied. The small background from $B \to D_s^{(*)} K \ell \nu$ decays is estimated from MC simulation using a simple phase space model. The backgrounds from continuum processes and $B \to D_s^{(*)} K \ell \nu$ decays are subtracted in each counting region from the yields measured in $\Upsilon(5S)$ data.

\subsection{Signal extraction}

\begin{figure*}
\centering
\subfigure[$D_s^-e^+$]{\includegraphics[width=0.24\textwidth]{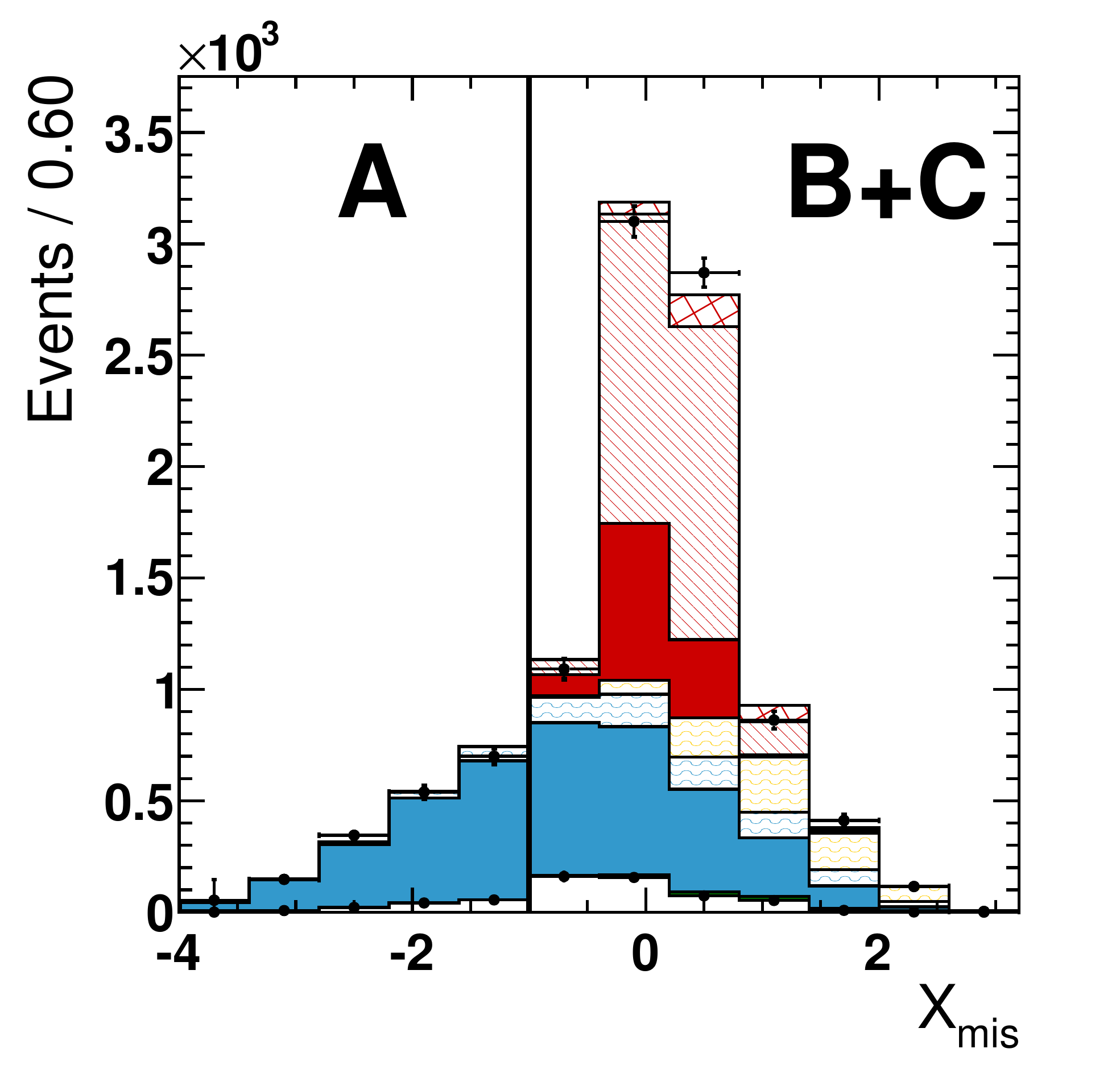}}
\subfigure[$D_s^-\mu^+$]{\includegraphics[width=0.24\textwidth]{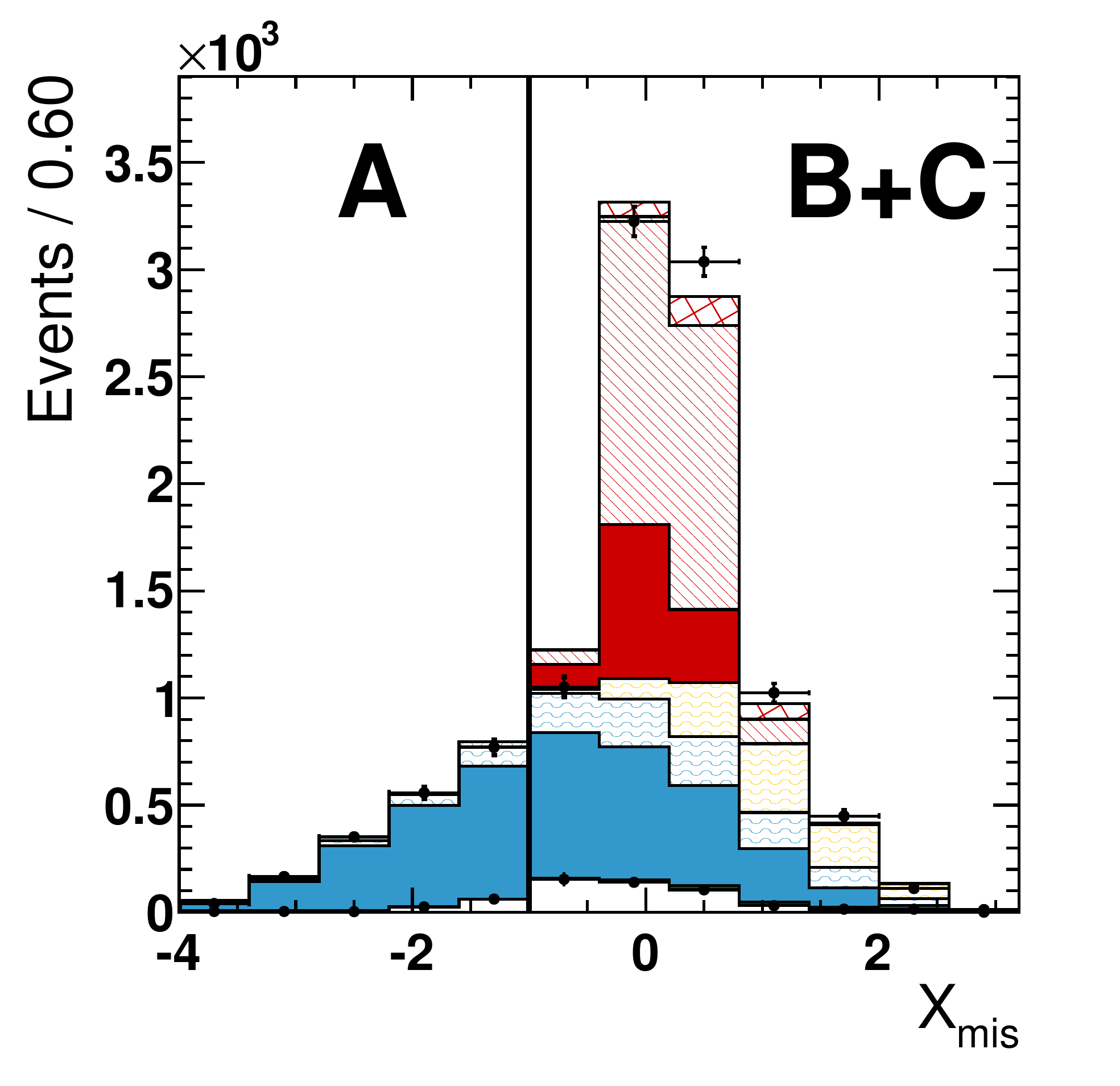}}
\subfigure[$D_s^{*-}e^+$]{\includegraphics[width=0.24\textwidth]{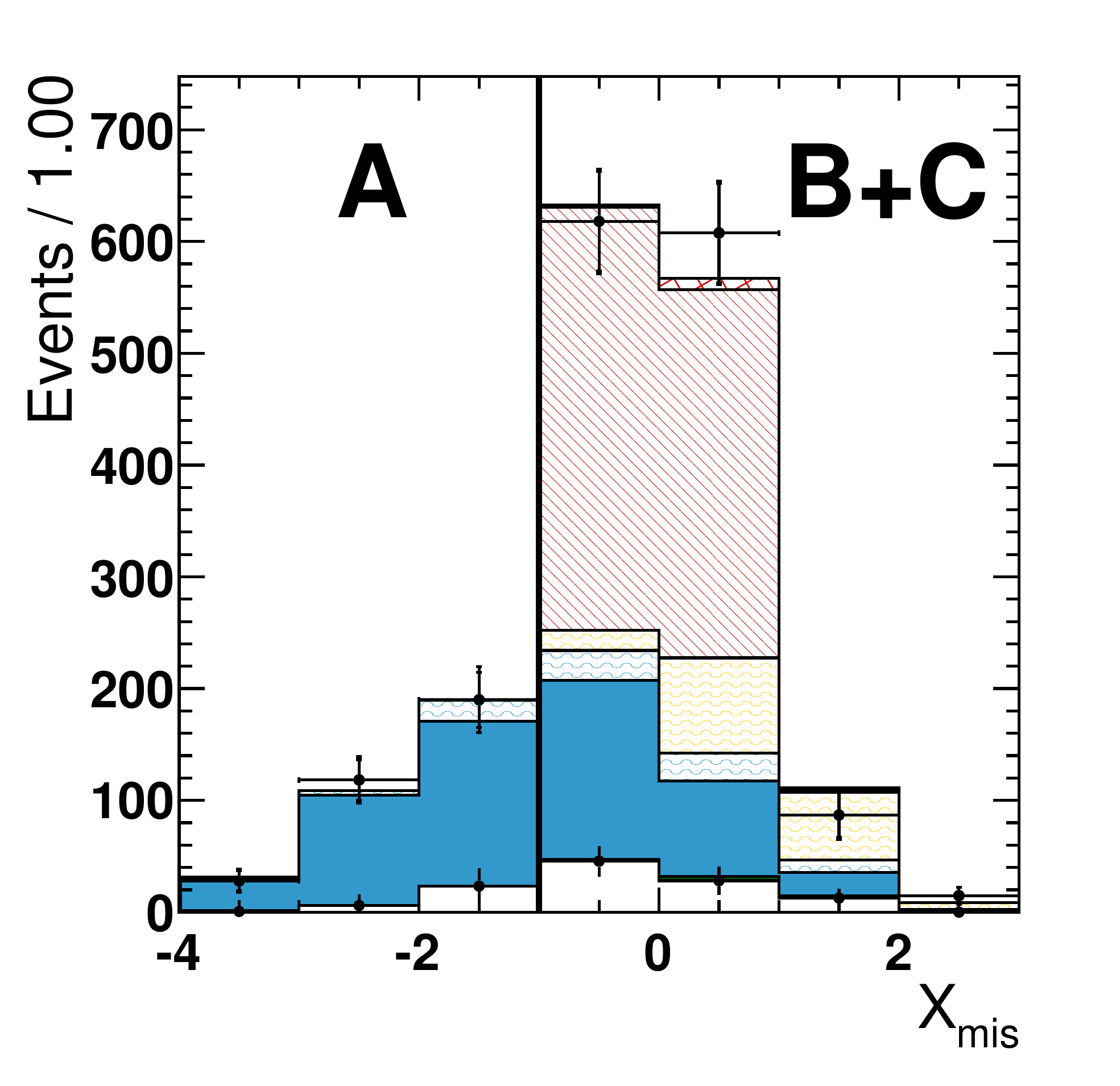}}
\subfigure[$D_s^{*-}\mu^+$]{\includegraphics[width=0.24\textwidth]{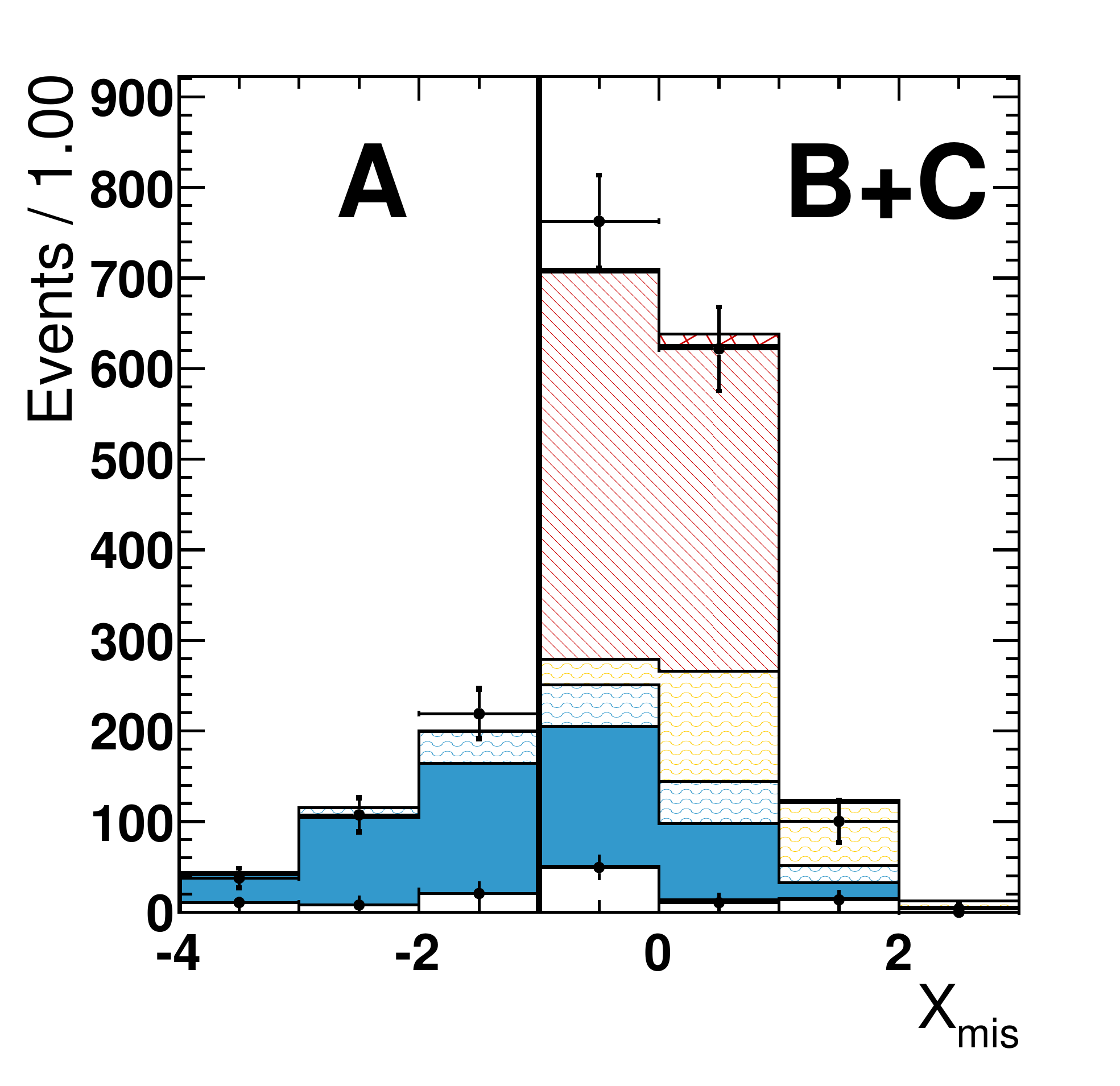}}
\subfigure[$D_s^-e^+$]{\includegraphics[width=0.24\textwidth]{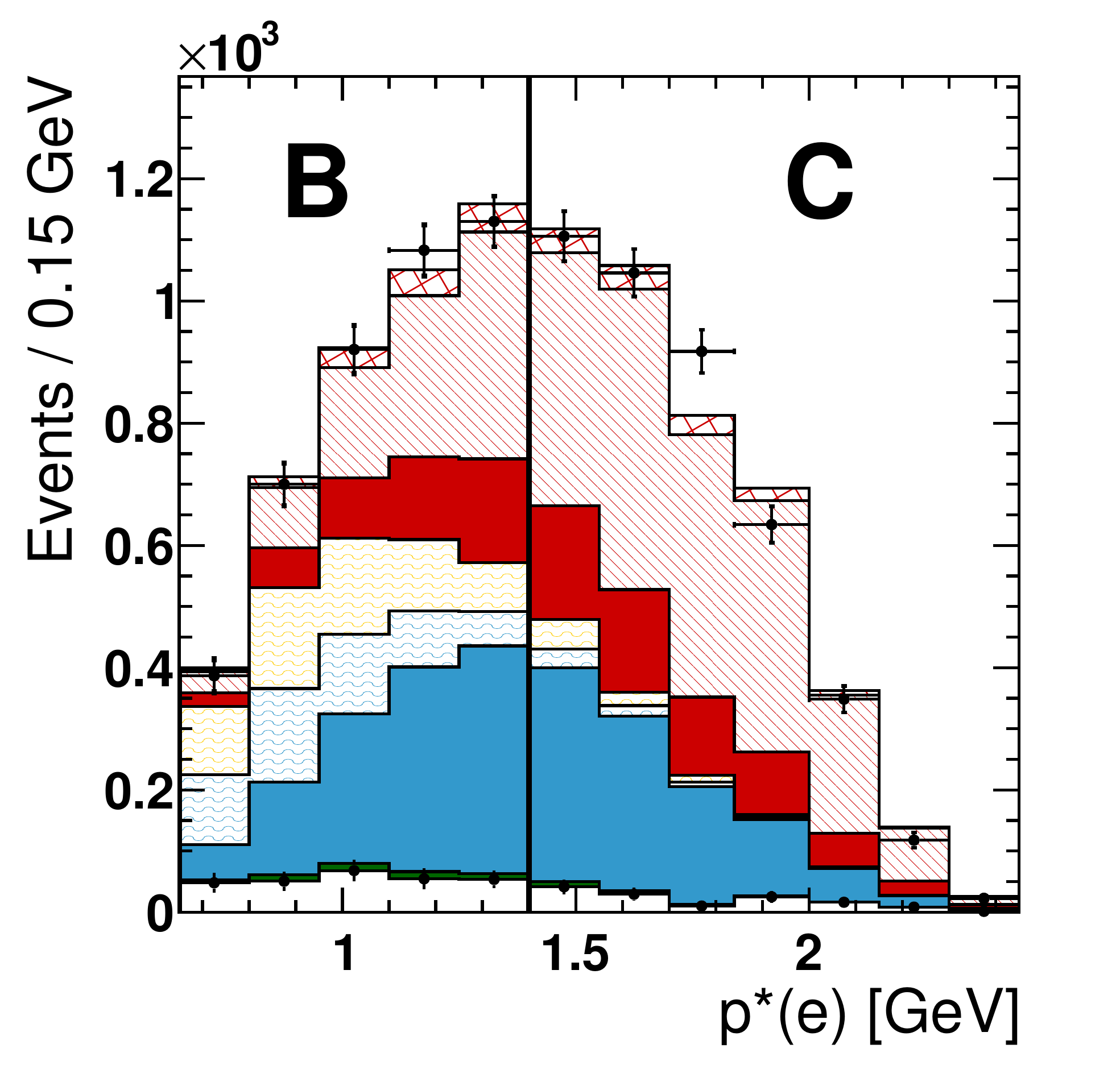}}
\subfigure[$D_s^-\mu^+$]{\includegraphics[width=0.24\textwidth]{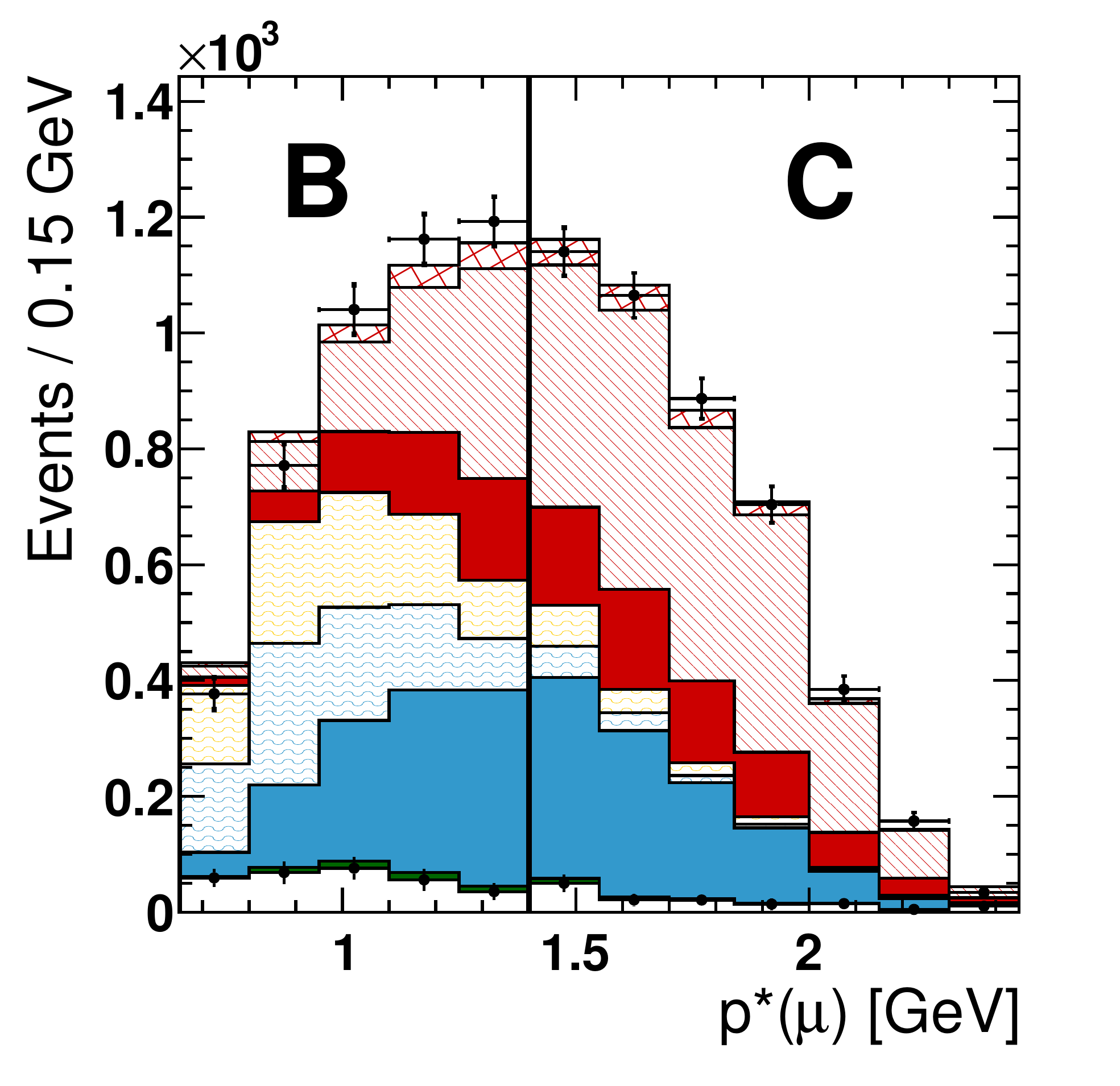}}
\subfigure[$D_s^{*-}e^+$]{\includegraphics[width=0.24\textwidth]{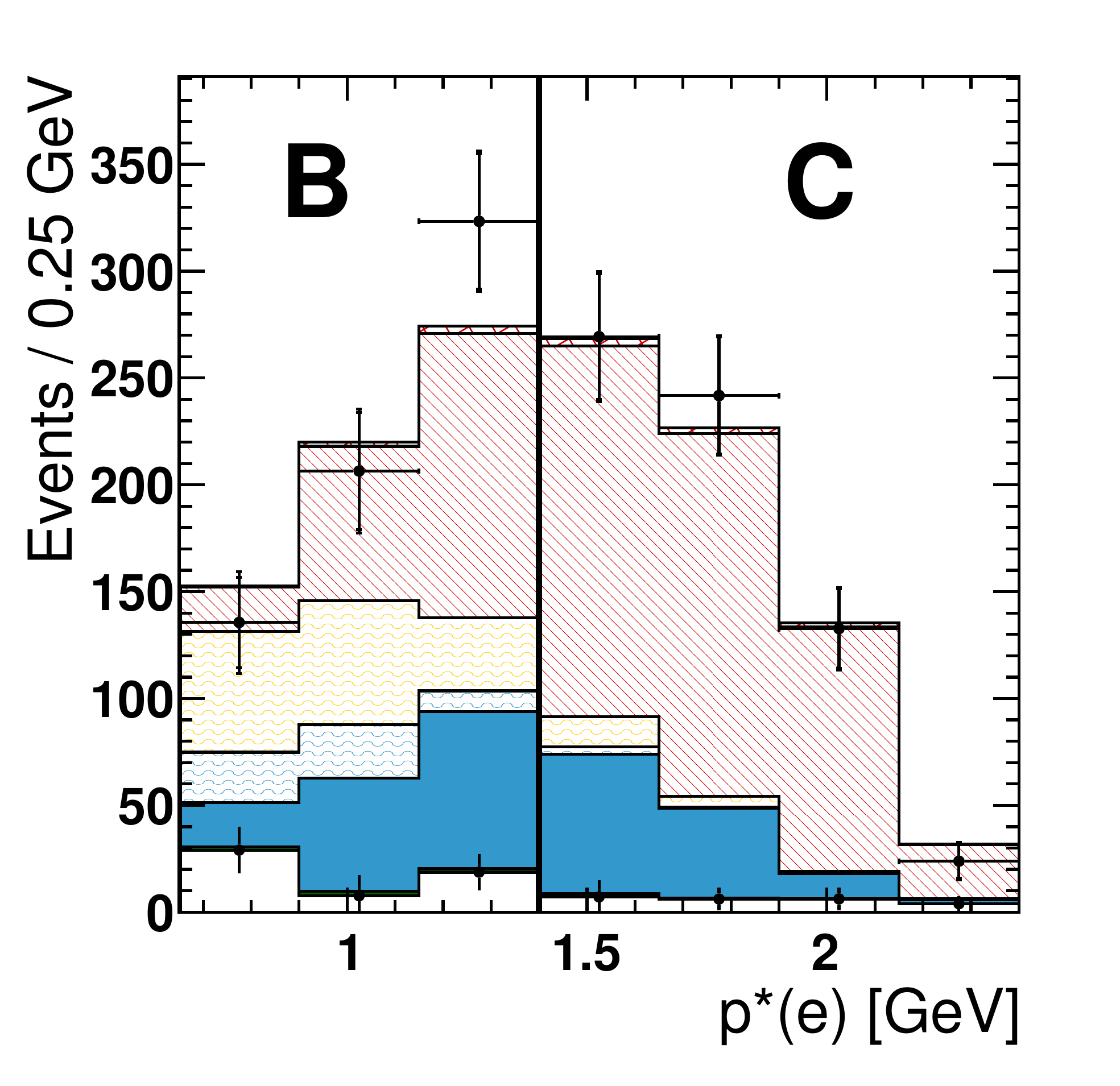}}
\subfigure[$D_s^{*-}\mu^+$]{\includegraphics[width=0.24\textwidth]{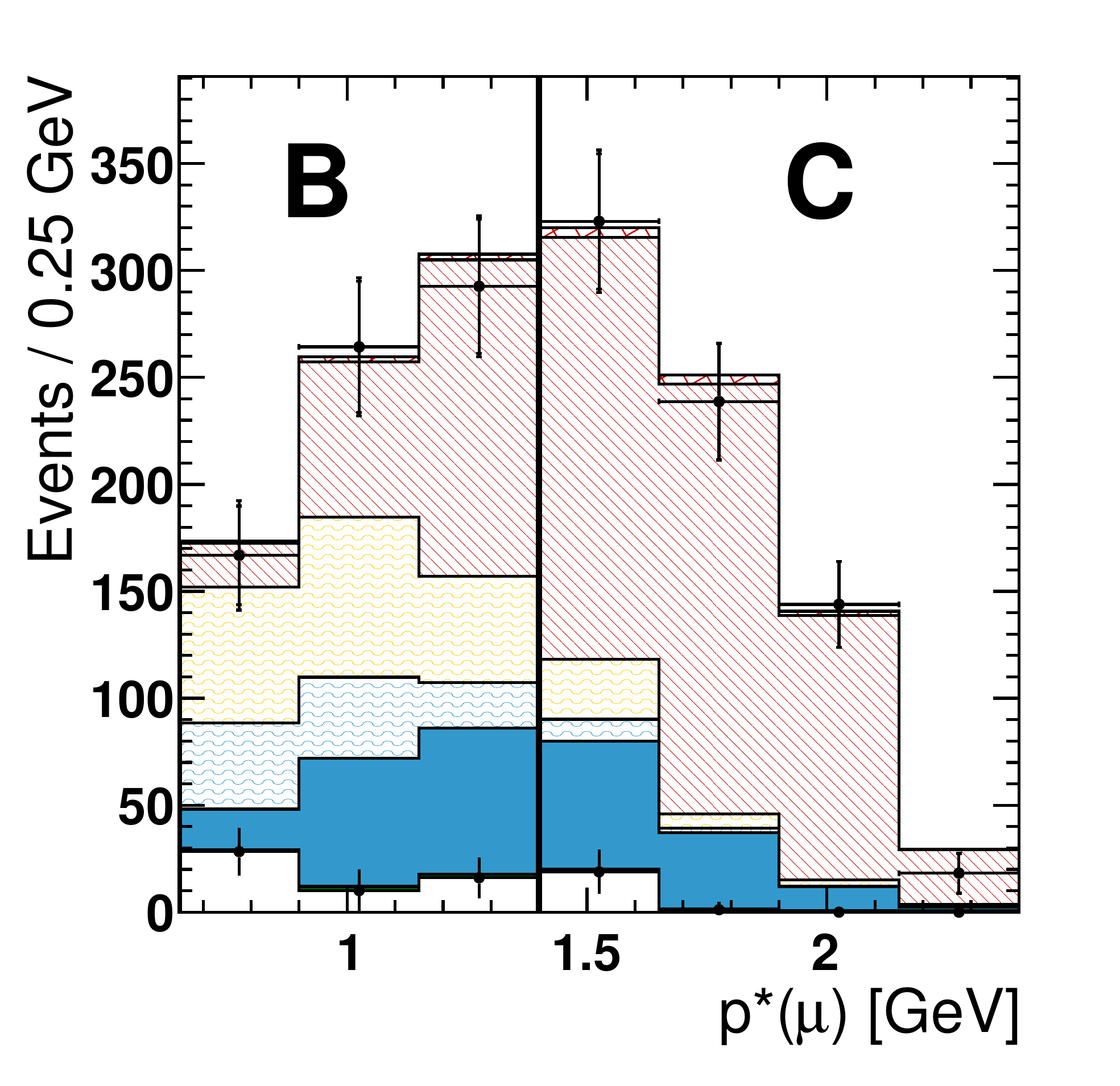}}
\caption{Distributions of $X_\text{mis}$ and $p^*(\ell)$ for reconstructed $D_s^{(*)-}\ell^+$ events. The black points with uncertainty bars show the $D_s^{(*)-}$ yields in the $\Upsilon(5S)$ data determined by fits to the $M_{KK\pi}$ distributions for  $D_s^{-}\ell^+$
and the $\Delta M$ distributions for  $D_s^{*-}\ell^+$. The stacked histograms represent the signal and background expectations after applying the scale factors $a_j$ (see Table~\ref{tab:integrated_yields}). The components are, from bottom to top: continuum background (white), $B \to D_s^{(*)} K \ell \nu$ background (dark green), \emph{opposite-$B_{(s)}$} primary leptons (solid blue), \emph{opposite-$B_{(s)}$} secondary leptons and misidentified hadrons (hatched blue), \emph{same-$B_{(s)}$} background (hatched yellow), \emph{signal} $B_s \to D_s \ell \nu$ (solid red), \emph{signal} $B_s \to D_s^* \ell \nu$ (hatched red), \emph{signal} $B_s \to D_s^{**} \ell \nu$ (cross-hatched red). The vertical black line illustrates the division of the counting regions. The displayed binning of the $X_\text{mis}$ and $p^*(\ell)$ distributions is used only to illustrate the data-MC agreement; the signal yield, $N_\text{sig}$, is extracted from the measured $D_s^{(*)-}$ yields in the three counting regions A, B and C listed in Tables~\ref{tab:ds_yields} and \ref{tab:dsST_yields}.}
\label{fig:lookback}
\end{figure*}

After subtraction of the continuum and the $B \to D_s^{(*)} K \ell \nu$ background components, the remaining yields contain three contributions: \emph{opposite-$B_{(s)}$} background, \emph{same-$B_{(s)}$} backgrounds and \emph{signal}. The three contributions are constrained by the event yields in the three counting regions. We introduce a scale factor, $a_j$, for each contribution, $j$. The determination of the scale factors is equivalent to solving a system of three linear equations with three unknowns. In order to obtain the uncertainties on the scale factors, we 
minimize:
\begin{equation}
\chi^2 = \sum_{i=A,B,C} ~
\frac{\left(D_i - \sum_j a_j N_{i,j}\right)^2}
{(\Delta D_i)^2 + \sum_j (a_j \Delta N_{i,j})^2}\,,
\label{eq:chi2}
\end{equation}
where the index $i$ runs over the three counting regions, $D_i$ is the event yield determined by the fits to the $M_{KK\pi}$ or $\Delta M$ distributions in data, and $\Delta D_i$ is the statistical uncertainty of these fits, $N_{i,j}$ is the MC prediction for the contribution $j$, and $\Delta N_{i,j}$ is its statistical uncertainty. Table \ref{tab:integrated_yields} lists the scale factors, $a_j$, obtained from the $\chi^2$ minimization and the signal yields,
\begin{equation}
N_\text{sig} = a_\text{sig} \cdot \sum_i N_{i,\text{sig}}\,.
\label{eq:sigyield}
\end{equation}
Figure~\ref{fig:lookback} shows the $X_\text{mis}$ and $p^*(\ell)$ distributions in the three counting regions after applying the scale factors $a_j$.

\begin{table*}
\centering
\caption{The scale factors, $a_j$, for the MC components obtained by minimizing the $\chi^2$ function defined in Eq.~(\ref{eq:chi2}). The errors are the statistical uncertainties of the data and the MC sample. The signal yields are determined from Eq.~(\ref{eq:sigyield}). The yields of the other components are given in Tables~\ref{tab:ds_yields} and~\ref{tab:dsST_yields}. The signal efficiencies are obtained by averaging over the efficiencies for the $D_s\ell\nu$, $D_s^*\ell\nu$ and $D_s^{**}\ell\nu$ modes, taking into account the expected relative abundance of the signal components. The given errors of the efficiencies are the statistical uncertainties of the MC sample.}
\begin{tabular}{l|ccc|ll}
\hline \hline
 & \multicolumn{3}{c|}{Scale factors}  \\
Channel & \emph{Opposite-$B_{(s)}$} & \emph{Same-$B_{(s)}$} & \emph{Signal} & Signal yield & Efficiency $[\%]$ \\ 
\hline
$D_s X e \nu$ & 1.02 $\pm$ 0.04 & 1.00 $\pm$ 0.20 & 1.06 $\pm$ 0.04 & 4470 $\pm$ 161 & 16.9 $\pm$ 0.1\\ 
$D_s X \mu \nu$ & 1.06 $\pm$ 0.04 & 0.94 $\pm$ 0.16 & 1.09 $\pm$ 0.04 & 4411 $\pm$ 161 & 16.3 $\pm$ 0.1 \\
$D_s^* X e \nu$ & 0.89 $\pm$ 0.12 & 1.66 $\pm$ 0.71 & 1.00 $\pm$ 0.11  & \hphantom{0}724 $\pm$ 79\hphantom{0} & \hphantom{0}4.6 $\pm$ 0.1 \\  
$D_s^* X \mu \nu$ & 0.96 $\pm$ 0.12 & 1.50 $\pm$ 0.58 & 1.13 $\pm$ 0.12 & \hphantom{0}804 $\pm$ 86\hphantom{0} & \hphantom{0}4.6 $\pm$ 0.1 \\ 
\hline \hline
\end{tabular}
\label{tab:integrated_yields}
\end{table*}

\begin{table*}
\centering
\caption{The $D_s^- \ell^+$ yields obtained from the $M_{KK\pi}$ fits to  $\Upsilon(5S)$ data in the three counting regions (A, B, C) and the corresponding signal and background expectations. The scale factors from Table~\ref{tab:integrated_yields} obtained by minimizing Eq.~(\ref{eq:chi2}) are applied to the MC expectations (3) -- (5). The errors are the statistical uncertainties of the data and MC samples, respectively, and do not contain the scale factor uncertainties. Uncertainties are omitted if they are smaller than $0.5$.}
\begin{tabular}{lrrrrrr}
\hline \hline
& \multicolumn{3}{c}{Electrons} & \multicolumn{3}{c}{Muons} \\
 & A \hphantom{00} & B \hphantom{00} & C \hphantom{00} & A \hphantom{00} & B \hphantom{00} & C \hphantom{00} \\
 \hline
$\Upsilon(5S)$ data & 1807 $\pm$ 53 & 4274 $\pm$ 87 & 4215 $\pm$ 82 & 1902 $\pm$ 54 & 4544 $\pm$ 89 & 4375 $\pm$ 81 \\ 
\hline
(1) Continuum (scaled off-resonance data) & 130 $\pm$ 34 & 278 $\pm$ 37 & 137 $\pm$ 22 & 102 $\pm$ 32 & 298 $\pm$ 40 & 134 $\pm$ 25 \\ 
(2) $B \to D_s K \ell \nu$ & 0 \hphantom{$\pm$} \hphantom{00} & 48 $\pm$ 7\hphantom{0} & 18 $\pm$ 4\hphantom{0} & 0 \hphantom{$\pm$} \hphantom{00} & 46 $\pm$ 7\hphantom{0} & 18 $\pm$ 4\hphantom{0} \\ 
(3) Opposite-$B_{(s)}$, secondary leptons, mis-ID hadrons & 110 $\pm$ 4\hphantom{0} & 555 $\pm$ 10 & 61 $\pm$ 3\hphantom{0} & 205 $\pm$ 6\hphantom{0} & 826 $\pm$ 12 & 107 $\pm$ 4\hphantom{0} \\ 
(3) Opposite-$B_{(s)}$, primary leptons & 1565 $\pm$ 16 & 1165 $\pm$ 14 & 1032 $\pm$ 13 & 1594 $\pm$ 17 & 1081 $\pm$ 14 & 1043 $\pm$ 14 \\ 
(4) Same-$B_{(s)}$ background & $0$ \hphantom{$\pm$ 00} &  638 $\pm$ 10 & 89 $\pm$ 4\hphantom{0} & 1 \hphantom{$\pm$ 00} & 798 $\pm$ 11 & 158 $\pm$ 5\hphantom{0} \\ 
(5) Signal ($D_s\ell\nu$) & 0 \hphantom{$\pm$ 00} & 492 $\pm$ 9\hphantom{0} & 669 $\pm$ 11 & 0 \hphantom{$\pm$ 00} &  489 $\pm$ 9\hphantom{0} & 693 $\pm$ 11 \\ 
(5) Signal ($D_s^*\ell\nu$) & 1 \hphantom{$\pm$ 00} & 951 $\pm$ 13 & 2072 $\pm$ 19 & 0 \hphantom{$\pm$ 00} & 872 $\pm$ 13 & 2072 $\pm$ 19 \\ 
(5) Signal ($D_s^{**}\ell\nu;~D_s^{**}\to D_s^*$) & 0 \hphantom{$\pm$ 00} & 28 $\pm$ 2\hphantom{0} & 41 $\pm$ 3\hphantom{0} & 0 \hphantom{$\pm$ 00} & 26 $\pm$ 2\hphantom{0} & 40 $\pm$ 3\hphantom{0} \\ 
(5) Signal ($D_s^{**}\ell\nu;~D_s^{**}\nrightarrow D_s^*$) & 0 \hphantom{$\pm$ 00} & 117 $\pm$ 5\hphantom{0} & 98 $\pm$ 4\hphantom{0}  & 0 \hphantom{$\pm$ 00} & 109 $\pm$ 4\hphantom{0} & 110 $\pm$ 4\hphantom{0}\\ 
\hline \hline
\end{tabular}
\label{tab:ds_yields}
\end{table*}

\begin{table*}
\centering
\caption{The $D_s^{*-} \ell^+$ yields obtained from the $\Delta M$ fits to  $\Upsilon(5S)$ data in the three counting regions (A, B, C) and the corresponding signal and background expectations. The scale factors from Table~\ref{tab:integrated_yields} obtained by minimizing Eq.~(\ref{eq:chi2}) are applied to the MC expectations (3) -- (5). The errors are the statistical uncertainties of the data and MC samples, respectively, and do not contain the scale factor uncertainties.}
\begin{tabular}{lrrrrrr}
\hline \hline
& \multicolumn{3}{c}{Electrons} & \multicolumn{3}{c}{Muons} \\
 &  A \hphantom{00} & B \hphantom{00} & C \hphantom{00} &  A \hphantom{00} & B \hphantom{00} & C \hphantom{00}\\
\hline
$\Upsilon(5S)$ data & 336 $\pm$ 33 & 656 $\pm$ 48 & 662 $\pm$ 46 & 370 $\pm$ 35 & 739 $\pm$ 52 & 741 $\pm$ 50\\ 
\hline
(1) Scaled off-resonance data &32 $\pm$ 22 & 61 $\pm$ 17 & 24 $\pm$ 11 & 49 $\pm$ 19 & 54 $\pm$ 18 & 20 $\pm$ 11 \\ 
(2) $B \to D_s K \ell \nu$ & 0 \hphantom{$\pm$ 00} & 6 $\pm$ 2\hphantom{0} & 2 $\pm$ 1\hphantom{0} & 0 \hphantom{$\pm$ 00} & 4 $\pm$ 2\hphantom{0} & 2 $\pm$ 1\hphantom{0} \\ 
(3) Opposite-$B_{(s)}$, secondary leptons, mis-ID hadrons & 24 $\pm$ 2\hphantom{0} & 60 $\pm$ 3\hphantom{0} & 4 $\pm$ 1\hphantom{0} & 48 $\pm$ 3\hphantom{0} & 99 $\pm$ 4\hphantom{0} & 13 $\pm$ 1\hphantom{0} \\ 
(3) Opposite-$B_{(s)}$, primary leptons & 279 $\pm$ 6\hphantom{0} & 147 $\pm$ 5\hphantom{0} & 120 $\pm$ 4\hphantom{0} & 273 $\pm$ 7\hphantom{0} & 147 $\pm$ 5\hphantom{0} & 109 $\pm$ 4\hphantom{0}\\ 
(4) Same-$B_{(s)}$ background & 0 \hphantom{$\pm$ 00} & 151 $\pm$ 6\hphantom{0} & 20 $\pm$ 2\hphantom{0} & 0 \hphantom{$\pm$ 00} & 188 $\pm$ 7\hphantom{0} & 39 $\pm$ 3\hphantom{0}\\ 
(5) Signal ($D_s^*\ell\nu$) & 0 \hphantom{$\pm$ 00} & 227 $\pm$ 6\hphantom{0} & 483 $\pm$ 9\hphantom{0} & 0 \hphantom{$\pm$ 00} & 241 $\pm$ 7\hphantom{0} & 547 $\pm$ 10\\ 
(5) Signal ($D_s^{**}\ell\nu;~D_s^{**}\to D_s^*$) & 0 \hphantom{$\pm$ 00} &  6 $\pm$ 1\hphantom{0} & 8 $\pm$ 1\hphantom{0} & 0 \hphantom{$\pm$ 00} & 6 $\pm$ 1\hphantom{0} & 11 $\pm$ 1\hphantom{0} \\ 
\hline \hline
\end{tabular}
\label{tab:dsST_yields}
\end{table*}

\section{Systematic uncertainties}

\begin{table}
\centering
\caption{Relative systematic uncertainties on the signal yields in~\%.}
\label{tab:allsystematics}
\begin{tabular}{lcccc}
\hline \hline
& $D_s X e \nu$ & $D_s X \mu \nu$ & $D_s^* X e \nu$ & $D_s^* X \mu \nu$ \\
\hline
\multicolumn{5}{l}{\bf{Detector}} \\
Tracking efficiency & 1.4 & 1.4 & 1.4 & 1.4 \\
Photon efficiency & --- & --- & 2.0 & 2.0 \\
Kaon and pion ID  & 1.4 & 1.4 & 1.4 & 1.4 \\
Lepton efficiency & 1.0 & 1.6 & 1.0 & 1.6 \\
Hadron misidentification & 0.1 & 1.3 & 0.1 & 1.9 \\
\hline
\multicolumn{5}{l}{\bf{Signal and background modeling}} \\
PDF for $M_{KK\pi}$ and $\Delta M$ fits  & 3.0 & 3.0 & 5.0 & 5.0  \\
Continuum shape & 1.2 & 0.3 & 1.2 & 0.3 \\
$B \to D_s^{(*)} K \ell \nu$ modeling & 0.3 & 0.3 & 0.1 & 0.1 \\
\multicolumn{5}{l}{\emph{Signal}} \\
\hspace{0.5cm} Composition & 4.8 & 4.8 & 0.3 & $<$ 0.1 \\
\hspace{0.5cm} Form factors & 0.9 & 1.0 & 1.0 & 1.0 \\
\hspace{0.5cm} Efficiency & 3.1 & 3.1 & 3.0 & 3.0 \\
\multicolumn{5}{l}{\emph{Opposite-$B_{(s)}$} background} \\
\hspace{0.5cm} Composition & 1.6 & 2.2 & 1.0 & 2.5 \\
\hspace{0.5cm} $B_s$ fraction & 0.2 & 0.2 & $<$ 0.1 & $<$ 0.1 \\
\hspace{0.5cm} Shape & 1.0 & 1.0 & 1.0 & 1.0 \\
\multicolumn{5}{l}{\emph{Same-$B_{(s)}$} background} \\
\hspace{0.5cm} Composition and shape & 0.3 & 0.3 & 0.4 & 0.7 \\
$B_s$ production mode & 0.1 & 0.1 & 0.3 & 0.3 \\
Beam energy & 1.0 & 1.0 & 0.5 & 0.5 \\
\hline
{\bf Total} & 7.3 & 7.6 & 6.9 & 7.6 \\
\hline \hline
\end{tabular}
\end{table}

The different sources of systematic uncertainties on the measured signal yields are described below. They comprise detector effects and the modeling of the signal and backgrounds. An overview can be found in Table~\ref{tab:allsystematics}.

\subsection{Detector effects}

The uncertainty on the track finding efficiency is 0.35\% per track and thus 1.4\% for four tracks. The photon efficiency is studied with radiative Bhabha events, from which the uncertainty is estimated to be 2\%. The calibration of kaon and pion identification efficiencies is estimated from a sample of reconstructed $D^{*+} \to D^0 \pi^+; D^0 \to K^- \pi^+$ decays. A variation of the obtained calibration factors within their uncertainties changes the measured signal yield by 1.4\%. The efficiency of the lepton identification is estimated using the two processes $\gamma\gamma \to \ell^+\ell^-$ and $J/\psi \to \ell^+\ell^-$. The corresponding uncertainties on the measured signal yields are 1.0\% and 1.6\% for the electron and muon modes, respectively. The rates of hadrons being misidentified as leptons  are estimated with the aforementioned $D^{*+}$ sample. The uncertainties due to this estimation are 0.1\% ($D_s X e \nu$), 1.3\% ($D_s X \mu \nu$), 0.1\% ($D_s^* X e \nu$) and 1.9\% ($D_s^* X \mu \nu$). 

\subsection{Signal and background modeling}

To study uncertainties of the PDFs in the $D_s$ and $D_s^*$ fits, we repeat the fits with alternative fit models and assign the resulting change of the signal yield as systematic uncertainty. Herein, we focus on the tails of the signal peaks because they can be easily assigned in the fit to the background component without deteriorating the agreement of the data with the fitted curve.  The signal PDF in the $D_s$ fits is modified by replacing the second Gaussian function by a bifurcated Gaussian function. This choice is motivated by a small asymmetry of the signal peak due to final state radiation. The normalization and the widths of the bifurcated Gaussian function are determined relative to the normalization and width of the Gaussian function from a fit to signal MC. These parameters are fixed in the fit to data. Based on the observed change of the signal yield, we assign a 3\% PDF uncertainty. In the $D_s^*$ fits, the tails of the signal peak are described by the Gaussian component of the signal PDF. When this Gaussian function is removed from the signal PDF, i.e. a Crystal Ball function only is used (cf. Ref.~\cite{DeltaMAlt}), the signal yields decrease by 5\%. Hence, we estimate the PDF uncertainty with 5\%.

The uncertainty due to the continuum scale factor, $S$, is negligible. The uncertainty due to the shape correction for the continuum background is estimated as the full difference of the result with and without the correction applied, which is 1.2\% and 0.3\% for electrons and muons, respectively. To estimate the influence of the choice of the $B \to D_s^{(*)} K \ell \nu$ decay model, we replace the phase space model used in the nominal result with the ISGW2 model~\cite{Scora:1995ty}, assuming that the decay proceeds via $B \to D_0^* \ell \nu;~ D_0^* \to D_s^{(*)} K$. The use of this alternative model increases the signal yields by 0.3\% and 0.1\% for the $D_s X \ell \nu$ and $D_s^* X \ell \nu$ channels, respectively. We also vary the $B \to D_s^{(*)} K \ell \nu$ branching fraction by the measured uncertainty and observe no significant change in the measured yield. We test the stability of the signal extraction when the boundary between counting region B and C is varied between $p^*_\ell = 1.3$ and $1.5\,{\rm GeV}$. The resulting change of the signal yields is consistent with the expected change due to the increase/decrease of statistics in the respective counting regions and, therefore, no systematic uncertainty is assigned.

The systematic uncertainty on the \emph{signal} composition in the $D_s X \ell\nu$ channels is obtained by evaluating the effect of scaling the relative amount of $B_s \to D_s^*\ell\nu$ decays up and down by 30\% and adjusting the $B_s \to D_s\ell\nu$ component such that the total number of MC events is conserved. This variation covers most of the recent theory predictions and causes a $4.7\%$ change of the  signal yields. To estimate the impact of $D_s^{**} \to D_s X$ crossfeed, we double the $B_s \to D_s^{**}\ell\nu$ contribution in the signal component, which increases the signal yield by 1\%. The $D_s^* X \ell \nu$ \emph{signal} component is expected to be dominated by $B_s \to D_s^*\ell\nu$ decays and hence the uncertainty due to the amount of $D_s^{**} \to D_s^{*} X$ crossfeed is negligible for this channel.

The $B \to D^{(*)}\ell\nu$ form factor parameters from Ref.~\cite{hfag} used to simulate the $B_s \to D_s^{(*)} \ell \nu$ decays are measured with an accuracy of 2-3\%. However, SU(3) flavor symmetry breaking effects may cause deviations at the order of 10\%~\cite{Blasi:1993fi}. To account for these differences, we vary each form factor parameter of a given decay independently up and down by 10\%. The resulting average deviation from the nominal signal yield is added linearly for each variation. The uncertainty of the LLSW model for $B_s \to D_s^{**}\ell\nu$ decays is evaluated by repeating the measurement with different sets of model parameters, as specified in Ref.~\cite{llsw}. The total systematic uncertainty due to form factor modeling is given by the quadratic sum of the uncertainties from all decay modes and does not exceed 1\%. 

The \emph{signal} efficiencies are studied in bins of three distributions: the lepton momentum, the $D_s$ momentum, and the angle between the reconstructed $D_s$ meson and the lepton in the CM system.  A re-calculation of the average efficiencies based on the observed data yields changes the signal by at most 3.1\%. 

The modeling of the \emph{opposite-$B_{(s)}$} component is studied in same-sign $D_s^+\ell^+$ control samples. The same-sign selection ensures that these samples contain only \emph{opposite-$B_{(s)}$} combinations. Compared to the $D_s^{(*)-}\ell^+$ samples, the relative contribution of $B_s$ decays is enhanced in this control sample. Two components of the \emph{opposite-$B_{(s)}$} sample are  distinguished: (1) primary leptons and (2) secondary leptons and hadron tracks misidentified as leptons. Scale factors for the normalisation of these two MC components are determined from fits to the $p^*(\ell)$ distributions of the $D_s^+\ell^+$ samples. The obtained scale factors are in agreement within the fit uncertainties of about 10\%. A variation of the normalisations of the two components in the $D_s^{(*)-}\ell^+$ samples within this 10\% uncertainty changes the signal yields between 1.0\% and 2.5\%, depending on the reconstructed channel. We also vary the fraction of $B_s$ decays in the \emph{opposite-$B_{(s)}$} component by 20\%, corresponding to the uncertainty of the $B_s$ production rate, $f_s$~\cite{PDBook}. The resulting change of the signal yields is less than $0.2\%$. The shape uncertainty of the \emph{opposite-$B_{(s)}$} component is evaluated in a data-driven way by using again the $D_s^+\ell^+$ samples, from which the event yields are determined in the three counting regions with the identical procedure as applied in the measurement. We calculate the ratios of data and MC yields for each counting region. These ratios range from 0.86 to 0.91 for electrons and from 0.96 to 0.97 for muons. We then modify the MC predictions for the \emph{opposite-$B_{(s)}$} component in the $D_s^-\ell^+$ simulation accordingly and study the impact on the measurement. The results change by less than 0.4\%, so an uncertainty of 1\% on the modeling of the \emph{opposite-$B_{(s)}$} component is a reasonable estimate, considering the differences between the $D_s^+\ell^+$ control samples and the $D_s^-\ell^+$ signal samples. The described approach cannot be transferred to the $D_s^{*-}\ell^+$ measurements because of the smaller sample sizes. However, the composition of the \emph{opposite-$B_{(s)}$} background in the $D_s^{*-}\ell^+$ sample is similar to the one in the $D_s^{-}\ell^+$ sample and hence the same uncertainty is assigned.

The decays contributing to the \emph{same-$B_{(s)}$} background component can be grouped into four classes with the corresponding fraction in the electron/muon channel given in parentheses: $B_{(s)} \to D_s^{(*)} X_c$ decays (70\% / 48\%), leptons stemming from $\tau$ produced via $B_s$ and $D_s$ decays (21\% / 16\%) and hadrons misidentified as leptons (9\% / 34\%). There are no significant differences in the composition between the $D_s^- \ell^+$ and the $D_s^{*-} \ell^+$ channels. We vary the fraction of leptons from $\tau$ decays and the fraction of misidentified hadrons by $\pm 50\%$ and take half the difference of the resulting signal yields as the systematic uncertainty, which is below 1\% for all measurements. Potential modeling uncertainties of the \emph{same-$B_{(s)}$} component are assumed to be covered by the large variation of the composition.

We estimate the impact of the uncertainty on the different $B_s$ production channels at the $\Upsilon(5S)$ energy by scaling the $B_s \bar{B}_s^*$ component up and down by 30\% and assign half of the change in the signal yield as the systematic uncertainty of 0.1\% and 0.3\% for the $D_s^- \ell^+$ and $D_s^{*-}\ell^+$ modes, respectively. The beam energy is conservatively varied by $\pm\unit[3]{MeV}$ and signal yield variations of 1\% and 0.5\% are observed for the $D_s^- \ell^+$ and $D_s^{*-}\ell^+$ modes, respectively. 

\section{Results and Discussion}

The semi-inclusive semileptonic $B_s$ branching fractions are calculated from Eq.~(\ref{eq:brafradef}). Since the $D_s \to K^+K^-\pi^+$ reconstruction mode is also used in the determination of $N_{B_s\bar{B}_s}$~\cite{NBs}, the $\mathcal{B}(D_s^+ \to K^+K^-\pi^+)$ branching fraction cancels out. Using the branching fraction ratio $\mathcal{B}(D_s^+ \to \phi\pi^+)/\mathcal{B}(D_s \to K^+K^-\pi^+) = (41.6 \pm 0.8 )\%$ and the branching fraction $\mathcal{B}(D_s^* \to D_s \gamma) = ( 94.2 \pm 0.7 )\%$~\cite{PDBook}, we obtain the semi-inclusive branching fractions:
\begin{eqnarray*}
D_s X e \nu~: & [8.1 \pm 0.3 (\text{stat}) \pm 0.6 (\text{syst}) \pm 1.4 (\text{ext})]\%\,, \\
D_s X \mu \nu~: & [8.3 \pm 0.3 (\text{stat}) \pm 0.6 (\text{syst}) \pm 1.5 (\text{ext})]\%\,, \\
D_s^* X e \nu~: & [5.2 \pm 0.6 (\text{stat}) \pm 0.4 (\text{syst}) \pm 0.9 (\text{ext})]\%\,, \\
D_s^* X \mu \nu~: & [5.7 \pm 0.6 (\text{stat}) \pm 0.4 (\text{syst}) \pm 1.0 (\text{ext})]\%\,. 
\end{eqnarray*}
The first uncertainty is the statistical uncertainty of the data and MC samples, the second is the systematic uncertainty of the measurement, and the last uncertainty is due to the external measurements of $N_{B_s\bar{B}_s}$ and $\mathcal{B}_{D_s^{(*)}}$. The electron and muon samples are statistically independent because only one candidate is selected per event. Taking into account that the systematic uncertainties are all correlated except the one for lepton identification, we calculate the combination of the measurements as weighted averages:
\begin{eqnarray*}
D_s X \ell \nu~ : & [8.2 \pm 0.2 (\text{stat}) \pm 0.6 (\text{syst}) \pm 1.4 (\text{ext})]\%\,, \\
D_s^* X \ell \nu~ : & [5.4 \pm 0.4 (\text{stat}) \pm 0.4 (\text{syst}) \pm 0.9 (\text{ext})]\%\,. 
\end{eqnarray*}

The obtained $B_s \to D_s X \ell \nu$ branching fraction can be compared to the difference between the inclusive branching fraction, $\mathcal{B}(B_s \to X_c \ell \nu)$, and the branching fraction of the $D_s^{**}\ell\nu$ modes, where the $D_s^{**}$ does not decay to a $D_s$ meson. The value of $\mathcal{B}(B_s \to X_c \ell \nu)$ is estimated to be $(10.0 \pm 0.4)\%$, using the branching fraction $\mathcal{B}(B^0 \to X_c \ell \nu)$~\cite{PDBook,Urquijo:2006wd}, an estimate for the ratio of the semileptonic widths of the $B_s$ and $B^0$ meson, $\Gamma_\text{sl}(B_s)/\Gamma_\text{sl}(B^0) = 0.99$~\cite{Bigi:2011gf} and the measured lifetimes of the $B^0$ and $B_s$ mesons~\cite{PDBook}. 

We assume that only the semileptonic decay modes with $D_{s1}(2536)$ and $D_{s2}(2573)$ mesons do not contain $D_s$ mesons in the final state. We obtain the estimate,
\begin{equation}
\begin{split}
& \mathcal{B}_\text{est}(B_s \to D_s X \ell \nu) \\
&= \mathcal{B}(B_s \to X_c \ell \nu) \cdot [ 1 - \\
&\hphantom{==} \mathcal{B}(B_s \to D_{s2} X \ell \nu) / \mathcal{B}(B_s \to X_c \ell \nu) - \\
&\hphantom{==} \mathcal{B}(B_s \to D_{s1} X \ell \nu) / \mathcal{B}(B_s \to X_c \ell \nu) ] \\ 
&=(9.1 \pm 0.4)\%\,,
\end{split}
\label{eq:brafraest}
\end{equation}
where the ratios $\mathcal{B}(B_s \to D_{s2} X \ell \nu) / \mathcal{B}(B_s \to X_c \ell \nu) = [3.3 \pm 1.0 (\text{stat}) \pm 0.4 (\text{syst})]\% $ and $\mathcal{B}(B_s \to D_{s1} X \ell \nu) / \mathcal{B}(B_s \to X_c \ell \nu) = [5.4 \pm 1.2 (\text{stat}) \pm 0.5 (\text{syst})]\%$ were measured at LHCb~\cite{LHCbCorr}.
The result of our measurement is in agreement with the estimate, $\mathcal{B}_\text{est}(B_s \to D_s X \ell \nu)$. The rate of $B_s \to D_s^{**} \ell \nu; D_s^{**} \to D_s^* X$ decays can be constrained from the comparison between the measured semi-inclusive branching fraction, $\mathcal{B}(B_s \to D_s^* X \ell \nu)$ with the exclusive theory predictions for $\mathcal{B}(B_s \to D_s^* \ell \nu)$. For example, using, the prediction from Ref.\cite{Faustov:2012mt}, one obtains $\mathcal{B}(B_s \to D_s^{**} \ell \nu; D_s^{**} \to D_s^* X)~<~2.0\%$ at the 90\% confidence level.

The measurement can also be used to determine $N_{B_s\bar{B}_s}$ using the estimate of the $B_s \to D_s X \ell \nu$ branching fraction from Eq.~(\ref{eq:brafraest}):
\begin{equation}
N_{B_s\bar{B}_s} = \frac{N_\text{sig}/[\epsilon \mathcal{B}(D_s \to \phi(K^+K^-) \pi^+)]}{2 \mathcal{B}_\text{est}(B_s \to D_s X \ell \nu)}\,.
\end{equation}
For $N_\text{sig}/\epsilon$, we insert the weighted average of the electron and muon modes, $N_\text{sig}/\epsilon = [26.7 \pm 0.7 (\text{stat}) \pm 2.0 (\text{syst})]~\times~10^3$; for $\mathcal{B}(D_s \to \phi(K^+K^-) \pi^+)$, we use the value $(2.24 \pm 0.10)\%$~\cite{PDBook}. We obtain $N_{B_s\bar{B}_s} = [6.53 \pm 0.17 (\text{stat}) \pm 0.49 (\text{syst}) \pm 0.41 (\text{ext})] \times 10^6$, corresponding to the cross section $\sigma(e^+e^- \to B_s^{(*)}\bar{B}_s^{(*)}) = [53.8 \pm 1.4 (\text{stat}) \pm 4.0 (\text{syst}) \pm 3.4 (\text{ext})]\,{\rm pb}$ at the CM energy $\sqrt{s} = 10.86\,{\rm GeV}$. The first two uncertainties are the statistical and systematic uncertainties from the measurement, respectively, and the last uncertainty is due to $\mathcal{B}(D_s \to \phi(K^+K^-) \pi^+)$ and $\mathcal{B}_\text{est}(B_s \to D_s X \ell \nu)$. The obtained result is in agreement with $N_{B_s\bar{B}_s} = (7.1 \pm 1.3) \times 10^6$ obtained by Belle with a different technique~\cite{NBs} and has a significantly improved precision.
 
\section{Summary}
We have presented the first measurements of the semi-inclusive branching fractions of  $B_s \to D_s X \ell \nu$ and $B_s \to D_s^* X \ell \nu$ decays. The measured branching fractions are $\mathcal{B}(B_s \to D_s X \ell \nu) = [8.2 \pm 0.2 (\text{stat}) \pm 0.6 (\text{syst}) \pm 1.4 (\text{ext})]\%$ and $\mathcal{B}(B_s \to D_s^* X \ell \nu) = [5.4 \pm 0.4 (\text{stat}) \pm 0.4 (\text{syst}) \pm 0.9 (\text{ext})]\%$. In addition, the analysis of these decays provides the currently most precise estimate of the $B_s^{(*)}\bar{B}_s^{(*)}$ production cross-section at the CM energy $\sqrt{s} = 10.86 {\rm GeV}$: $\sigma(e^+e^- \to B_s^{(*)}\bar{B}_s^{(*)}) = (53.8 \pm 1.4 (\text{stat}) \pm 4.0 (\text{syst}) \pm 3.4 (\text{ext})) {\rm pb}$.

\section{Acknowledgements}
We thank the KEKB group for the excellent operation of the
accelerator; the KEK cryogenics group for the efficient
operation of the solenoid; and the KEK computer group,
the National Institute of Informatics, and the 
PNNL/EMSL computing group for valuable computing
and SINET4 network support.  We acknowledge support from
the Ministry of Education, Culture, Sports, Science, and
Technology (MEXT) of Japan, the Japan Society for the 
Promotion of Science (JSPS), and the Tau-Lepton Physics 
Research Center of Nagoya University; 
the Australian Research Council and the Australian 
Department of Industry, Innovation, Science and Research;
Austrian Science Fund under Grant No.~P 22742-N16 and P 26794-N20;
the National Natural Science Foundation of China under Contracts 
No.~10575109, No.~10775142, No.~10875115, No.~11175187, and  No.~11475187; 
the Ministry of Education, Youth and Sports of the Czech
Republic under Contract No.~LG14034;
the Carl Zeiss Foundation, the Deutsche Forschungsgemeinschaft
and the VolkswagenStiftung;
the Department of Science and Technology of India; 
the Istituto Nazionale di Fisica Nucleare of Italy; 
National Research Foundation (NRF) of Korea Grants
No.~2011-0029457, No.~2012-0008143, No.~2012R1A1A2008330, 
No.~2013R1A1A3007772, No.~2014R1A2A2A01005286, No.~2014R1A2A2A01002734, 
No.~2014R1A1A2006456;
the Basic Research Lab program under NRF Grant No.~KRF-2011-0020333, 
No.~KRF-2011-0021196, Center for Korean J-PARC Users, No.~NRF-2013K1A3A7A06056592; 
the Brain Korea 21-Plus program and the Global Science Experimental Data 
Hub Center of the Korea Institute of Science and Technology Information;
the Polish Ministry of Science and Higher Education and 
the National Science Center;
the Ministry of Education and Science of the Russian Federation and
the Russian Foundation for Basic Research;
the Slovenian Research Agency;
the Basque Foundation for Science (IKERBASQUE) and 
the Euskal Herriko Unibertsitatea (UPV/EHU) under program UFI 11/55 (Spain);
the Swiss National Science Foundation; the National Science Council
and the Ministry of Education of Taiwan; and the U.S.\
Department of Energy and the National Science Foundation.
This work is supported by a Grant-in-Aid from MEXT for 
Science Research in a Priority Area (``New Development of 
Flavor Physics'') and from JSPS for Creative Scientific 
Research (``Evolution of Tau-lepton Physics'').


\begin{thebibliography}{99}
\bibitem{speedoflight}
Throughout this paper, the convention $c=1$ is used.
\bibitem{CC}
Throughout this paper, the inclusion of the charge conjugate mode decay is implied unless otherwise stated.
\end{thebibliography}
\end{document}